\documentclass[%
twocolumn,
showpacs,preprintnumbers,
 amsmath,amssymb,
 aps,
 superscriptaddress,
pra,
floatfix,
]{revtex4-1}

\usepackage{graphicx}
\usepackage{dcolumn}
\usepackage{bm}
\usepackage{hyperref}
\usepackage[utf8]{inputenc}
\usepackage{amsmath}
\usepackage[amssymb]{SIunits}
\usepackage{amsfonts}
\usepackage{amssymb}
\usepackage{subfigure}
\usepackage{mathtools}
\usepackage{dsfont}
\usepackage{color}
\usepackage{pstricks}
\usepackage{isotope}
\usepackage{calc}
\usepackage{braket}
\begin{document}

%\preprint{APS/123-QED}

\title{Cavity QED with an ultracold ensemble on a chip: prospects for strong magnetic
coupling at finite temperatures}

\author{Kathrin Henschel}
\affiliation{Institute for Theoretical Physics, Universität Innsbruck, Technikerstrasse 25, 6020 Innsbruck, Austria}

\author{Johannes Majer}
\author{Jörg Schmiedmayer}
\affiliation{
Atominstitut, TU-Wien,
Stadionallee 2,
1020 Vienna, Austria}

\author{Helmut Ritsch}
\affiliation{Institute for Theoretical Physics, Universität Innsbruck, Technikerstrasse 25, 6020 Innsbruck, Austria}

\date{\today}

\begin{abstract}
We study the nonlinear dynamics of an ensemble of cold trapped atoms with a hyperfine
transition magnetically coupled to a resonant microwave cavity mode. Despite the minute
single atom coupling one obtains strong coupling between collective hyperfine qubits and
microwave photons enabling coherent transfer of an excitation between the long lived
atomic qubit state and the mode. Evidence of strong coupling can be obtained from the
cavity transmission spectrum even at finite thermal photon number. The system makes it
possible to
study further prominent collective phenomena such as superradiant decay of an inverted
ensemble or the building of a narrowband stripline micromaser locked to an atomic hyperfine
transition.

\pacs{37.30.+i, 42.50.Pq}

\end{abstract}

%\keywords{Suggested keywords}%Use showkeys class option if keyword
                              %display desired
\maketitle

%\tableofcontents

\section{Introduction}
The idea of resonant coupling of an ensemble of atoms to a single cavity mode has been
addressed in numerous aspects and contexts, some dating back several
decades~\citep{tavis1968exact}.  Recently, in the context of quantum information
processing, such Hamiltonians attracted renewed attention because the ensemble can serve as
quantum memory with long coherence
times~\citep{duan2001long,rabl2006hybrid,imamoglu2008cqb,verdu:043603}.  Despite small
coupling of individual atoms, the strong collective coupling of the ensemble to a
particular cavity mode allows for the coherent transfer of an excitation to the ensemble,
its storage and its retrieval after some time shorter than the coherence time of the
system. Hence, due to collective effects, one can utilize atomic transitions and geometries
for which the strong coupling regime would not be accessible otherwise.

As a particularly striking example, one can even envisage the use of states that are very
weakly coupled to the field, for example, an optically forbidden hyperfine transition, which
only couple to the field via magnetic dipole interaction. What makes this idea attractive
and possibly feasible with current technology is the fact that it should be possible to
fabricate high-$Q$ stripline waveguide cavities on the superconducting surface of a
microchip, which confine the microwave mode to a very small effective volume and to
simultaneously trap a large ensemble of cold atoms very close to the surface. The
combination of high-$Q$ stripline waveguide cavities and atom-trapping technology surely
will involve new challenges, but there seem to be no fundamental problems. As already
demonstrated, such a cavity can be strongly coupled to on-chip Cooper-pair box
qubits~\citep{wallraff2004strong}. By
combining the two systems, one thus could establish a connection between the atomic
ensemble and solid-state qubits. This setup hence bridges an enormous range of time scales
starting from the sub-microsecond scale of solid-state qubits, over the millisecond lifetime of
microwave photons, to the atomic hyperfine coherence lifetime of seconds.

In the particular setup discussed here, the ensemble consists of a cloud of ultracold
\isotope[87]{Rb} trapped in an on-chip magnetic wire trap and pumped to one of the
trappable hyperfine levels, for example, $F=1$, $m_{F}=-1$.  The interaction between
the atoms and the field is dominated by the magnetic dipole transitions between
$\ket{F=1,m_{F}}$ and $\ket{F=2,m^{\prime}_{F}}$. These transitions are widely used for
hyperfine manipulations of cold atomic ensembles by externally injected
microwaves~\citep{bohi2009coherent}. We
assume in the following that the experimental setup guarantees that the cavity is
resonant with only one of the possible transitions (e.g. $m_{F}=-1$ and $m^{\prime}_{F}=1$
with transition frequency $\omega_{a}/\left( 2 \pi \right)=\unit{6.83}{\giga\hertz}$, corresponding to
$T\approx \unit{330}{\milli\kelvin}$), and hence allows for the atoms to be treated
as two-level systems. Actually, in some cases it is more favorable to use Raman-type
coupling employing an extra radio-wave field to choose a suitable microwave
transition~\citep{matthews1998dynamical}.

We ignore some of these technical details at this point and focus on the three main
topics: After the introduction of the model in Sec.~\ref{sec:model}, we first investigate conditions for strong coupling between the ensemble and the cavity
and the experimental consequences when one adds the obscuring effects of thermal photons
due to a finite cavity temperature. In Secs.~\ref{sec:num_small} and
~\ref{sec:cumulantexpansion} we discuss the methods we use, whereas in
Secs.~\ref{sec:Spectrum}--~\ref{sec:spec_driven} we address several
aspects of the resulting dynamics. Here the optically aligned ensemble, which has much
lower effective temperature, can be expected to act as a heat sink for the cavity mode
removing thermal photons. As the upper and lower hyperfine
states have a virtually infinite lifetime compared with other system time scales, we can
also completely invert the system, mimicking an effective negative temperature, and use it
to pump energy into the system. As a prominent example, we study in
Sec.~\ref{sec:superradiance} the superradiant decay of a fully inverted ensemble again
with some thermal photons initially present. Finally, we exhibit in Sec.~\ref{sec:laser}
the possibility of building an ultranarrow linewidth single-chip stripline micromaser
operating directly on an atomic clock type transition, which is in close analogy to an
optical-lattice-based setup, as recently suggested in~\citep{meiser2009pml}.

\section{Model}\label{sec:model}

\subsection{Collective atom-field Hamiltonian }
A single atom, formally represented here by a two-level system resonantly coupled to a
cavity mode, can be well described by the Jaynes-Cummings Hamiltonian. For $N$ two-level
systems trapped so close to each other in the cavity that they see the same field and
thus are coupled to the mode with equal strength $g$, we then get the generalized
Hamiltonian:
\begin{align}
	\operatorname{H}=\hbar \omega_{m} a^{\dagger }a + \frac{\hbar \omega_a}{2}\sum_j
        \sigma^{z}_j+\hbar g\sum_j \left( \sigma^{+}_j a+ a^{\dagger }
        \sigma^{-}_j \right)\ .
				\label{eq:hamiltonian_full}
\end{align}
with $a$ being the annihilation operator for a cavity photon, $\sigma^{+}_{j}$ being the
excitation operator for the $j$th two-level system and $\left[
\sigma_{i}^{+},\sigma_{j}^{-} \right]=\sigma_{i}^{z} \delta_{ij}$. The frequency of the
two-level systems and the mode are denoted by $\omega_{a}$ and $\omega_{m}$, respectively.
The coupling strength $g=\vec{B}\left( \vec{r} \right)\cdot\vec{\mu}/\hbar$ depends on the
strength of the magnetic field per photon $\vec{B}$ at the position $\vec{r}$ of the atoms
and the magnetic moment $\vec{\mu}$ of the considered transition.

What make an ensemble of atoms coupled to a cavity interesting are collective effects
emerging from the common coupling of all atoms to the same mode. This can be well
illustrated by introducing collective atomic operators
$\operatorname{S}^{\pm}=\sum_j\sigma_{j}^{\pm}$ and
$\operatorname{S}^{z}=\frac{1}{2}\sum_j\sigma_{j}^{z}$. The treatment in terms of
collective operators provides a convenient basis for classifying the possible states of the
ensemble and is therefore discussed here. As we see in Sec.~\ref{sec:noise}, we have
to resort to Hamiltonian~\eqref{eq:hamiltonian_full} in our particular treatment. 
The introduction of $\operatorname{S}^{\pm}$ and
$\operatorname{S}^{z}$ leads to the Tavis-Cummings form of this Hamiltonian:
\begin{align}
	\operatorname{H}_{\text{TC}}=\hbar \omega_{m} a^{\dagger }a + \hbar \omega_a
	\operatorname{S}^{z}+\hbar g \left( \operatorname{S}^{+} a+ a^{\dagger }
	\operatorname{S}^{-} \right)\,
   \label{eq:tavis}
\end{align}
where single photons are coupled to distributed (delocalized) excitations in the ensemble~\citep{tavis1968exact}. 
Let us shortly review some of its most known properties here.  Mathematically, the
collective operators follow the standard commutation relations for angular momentum
operators $\operatorname{\boldsymbol{S}}=\left( \operatorname{S}^{x},
\operatorname{S}^{y},\operatorname{S}^{z}\right)$, with $\operatorname{S}^{\pm}=\left(
\operatorname{S}^{x}\pm \mathrm{i}\operatorname{S}^{y}\right)$.
The corresponding eigenstates of $\operatorname{\boldsymbol{S}}^{2}$ and
$\operatorname{S}^{z}$ are the so-called Dicke
states $\ket{J,M}$, with $\operatorname{\boldsymbol{S}}^{2}\ket{J,M}=J(J+1)\ket{J,M}$ and
$\operatorname{S}^{z}\ket{J,M}=M\ket{J,M}$, where $J=0,1,\dots,N/2$ and $M=-J,\dots, J$.
Formally, a fully inverted ensemble corresponds to the maximum angular
momentum of $J=N/2$~\cite{haroche2006exploring, dicke1954coherence} and projection $M=N/2$.
Repeated application of the collective downward ladder operator $\operatorname{S}^{-}$
on the initial state $\ket{J,J}\mathrel{\widehat{=}}\ket{e,e,\dots,e}$ gives the lowest
state $\ket{J,-J}\mathrel{\widehat{=}}\ket{g,g,\dots,g}$.

The states in between are generated according to
\begin{align}
	\operatorname{S}^{\pm}\ket{J,M}=\sqrt{\left( J\pm M+1\right)\left( J\mp M
	\right)}\ket{J,M\pm1}\ .
	\label{eq:pm_norm}
\end{align}

The interaction can then be conveniently rewritten in terms of normalized collective operators
$\operatorname{\widetilde{S}}^{\pm}=\frac{1}{\sqrt{N}}\sum_i\sigma_{i}^{\pm}$ to obtain
\begin{align}
	\operatorname{\widetilde{H}}=\hbar \omega_{m} a^{\dagger }a + \hbar \omega_a
	\operatorname{\widetilde{S}}^{+}\operatorname{\widetilde{S}}^{-}+\hbar g_{\text{eff}} \left( \operatorname{\widetilde{S}}^{+} a+ a^{\dagger }
	\operatorname{\widetilde{S}}^{-} \right)\ ,
   \label{eq:tavis_particular}
\end{align}
with $g_{\text{eff}}=g\sqrt{N}$. Note that in the case where the atoms in the ensemble couple to the
cavity with different coupling constants $g_{i}$, we generalize to
$\operatorname{\widetilde{S}}^{\pm}=\frac{1}{g_{\text{eff}}}\sum_i g_{i}\sigma_{i}^{\pm}$,
with
$g_{\text{eff}}=\sqrt{\sum_{i}g_{i}^{2}}$. This reduces to $g_{\text{eff}}=g\sqrt{N}$ if
all $g_{i}$ are equal. To simplify matters, we remain with the case of equal coupling
strength. 

Allowing only one excitation in the system, we see that the ground state
$\ket{0}_{a}\mathrel{\widehat{=}}\ket{J,-J}\mathrel{\widehat{=}}\ket{g,g,\dots,g}$ is only
coupled to the symmetric atomic excitation state
$\operatorname{\widetilde{S}}^{+}\ket{0}_{a}=\ket{1}_{a}=\ket{J,-J+1}\mathrel{\widehat{=}}\frac{1}{\sqrt{N}}\left(\ket{e,g,g,\dots,g}+\ket{g,e,g,\dots,g}+\dots+\ket{g,\dots,g,e}\right)$,
while other atomic states with only one excitation play no role. Hence, in this form we
end up again with a two-level atomic system, where the dependence of the atom-cavity
coupling on the number of atoms is explicitly visible. Even for transitions with a very
small coupling constant $g$, strong coupling can be achieved for sufficiently large $N$.

Note that in Eq.~\eqref{eq:tavis_particular} we use
$\operatorname{\widetilde{S}}^{z}\approx-\frac{1}{2}+\frac{\operatorname{\widetilde{S}}^{+}\operatorname{\widetilde{S}}^{-}}{N}$,
where $\operatorname{\widetilde{S}}^{z}=\frac{1}{2N}\sum_{i}\sigma_{i}^{z}$.
This approximation is exactly valid only either for a single atom or the special case where we consider
only one excitation in the system. The constant $-\frac{1}{2}$ is neglected in the
Hamiltonian. In general, we find for a state with $J=\frac{N}{2}$ and $M=-J+s$
\begin{align}
	\bra{J,-J+s}\operatorname{\widetilde{S}}^{z}  \ket{J,-J+s}=
	-\frac{1}{2}+\frac{s}{2J}
	\label{eq:Szexp}
\end{align}
and
\begin{align}
	\bra{J,-J+s}\operatorname{\widetilde{S}}^{+}\operatorname{\widetilde{S}}^{-} \ket{J,-J+s}=
	s-\frac{s\left( s-1 \right)}{2J}\ .
	\label{eq:Spmexp}
\end{align}
For $s\ll N$ we neglect the second term on the right-hand side of Eq.~\eqref{eq:Spmexp}
and find the approximation for $\operatorname{\widetilde{S}}^{z}$, which becomes exact for
$s=1$. 

For large ensembles with few excitations this approximation is closely related to the
bosonization procedure. For $M\approx-J$ with $J=N/2$ and from
\begin{eqnarray}
	\left[ \operatorname{\widetilde{S}}^{+},\operatorname{\widetilde{S}}^{-}
	\right]&=&\frac{1}{N}\left[ \operatorname{S}^{+},\operatorname{S}^{-}
	\right]\notag\\&=&\frac{2\operatorname{S}^{z}}{N}=\left(-1+\mathcal{O}\left(
	\frac{1}{N}
	\right)\right)\mathds{1}\ ,
	\label{eq:commutations}
\end{eqnarray}
we find that for few excitations it is possible to identify
$\operatorname{\widetilde{S}}^{+}$ and $\operatorname{\widetilde{S}}^{-}$ with bosonic
creation and annihilation operators. Hence, we end up with a system of coupled oscillators,
for which a great deal of solution techniques exist.\par

Let us now come back to the atom-field interaction [Eq.~\eqref{eq:tavis_particular}]. It is
well known that the eigenstates are coherent superpositions of the two previously introduced
basis states, where the excitation is located either in the mode or in the ensemble.
Let $\ket{0}_{m}$ and $\ket{1}_{m}=a^{\dagger}\ket{0}_{m}$ be the possible states of the
mode and $\ket{0}_{a}$ and $\ket{1}_{a}$ be the ensemble states. With
$\omega_{a}=\omega_{m}$, the two eigenstates then read
\begin{align}
	\ket{+}&=\frac{1}{\sqrt{2}}\left( \ket{1}_{a}\ket{0}_{m}+\ket{0}_{a}\ket{1}_{m}
	\right)\ ,\\
	\ket{-}&=\frac{1}{\sqrt{2}}\left( \ket{1}_{a}\ket{0}_{m}-\ket{0}_{a}\ket{1}_{m}
	\right)\ ,
	\label{eq:eigenstates}
\end{align}
and as expected are separated by the energy difference $2g_{\text{eff}}$. Of course, the
system possesses more states containing essentially one excitation quantum, but those are
not directly coupled to the ground state if we consider only collective operators. The
collective operators couple states within one $J$ manifold, like the previously discussed
manifold with maximum angular momentum $J=N/2$ and $M=-J\dots J$. Taking into account
the manifolds of states with $J<N/2$, one can see that in general there is a large number
of states describing an ensemble with $n$ excitations. In the forthcoming calculations including
spontaneous emissions, such states with $J<N/2$ can be populated as well~\citep{haake1993superradiant}.
In addition, we also do not restrict the dynamics to a single excitation.

\subsection{Master equation including decoherence and thermal noise}\label{sec:noise}

In any realistic implementation of the preceding model, coupling of the thermal
environment to the field mode and the atoms is unavoidable.  This generates several
sources of noise and decoherence we have to address to be able to reliably describe
the dynamics. Despite its high $Q$ value, the microwave resonator still has a
non-negligible finite linewidth $\kappa=1/\tau$. In other words, a stored photon is likely
to be lost from the cavity after the time $\tau$. Similarly, atomic excitations are assumed
to decay with a rate that is, fortunately, in our case negligibly small
in practice~\citep{kasch2009cold}. However, we have to consider trap loss of atoms leaving the cavity mode,
which generates an effectively faster decay of the atomic excitation, denoted by the rate
$\gamma_{a}$. This can be to some
extent controlled by a suitable choice of the trapping states and trap geometry. An
additional and in general quite serious source of noise are thermal photons that leak into
the cavity. For an unperturbed cavity mode they lead to an average occupation number of
\begin{align}
   \bar{n}(\omega_{m},T)=\frac{\mathrm{e}^{-\frac{\hbar \omega_{m}}{k_B
	 T}}}{1-\mathrm{e}^{-\frac{\hbar\omega_{m}}{k_B T}}}\ ,
\end{align}
where $T$ denotes the temperature of the environment. In principle, such thermal photons
are also present on the atomic transition and lead to a thermalization of the optically
pumped atomic ensemble. Fortunately, the weak dipole moment of the atom renders this
thermalization rate so slow that it can be ignored at the experimentally relevant time
scales. In principle, even this rate could be collectively enhanced, but it largely addresses
collective states only very weakly coupled to the cavity mode.

Putting all these noise sources together, we can use standard quantum optical methods to
derive a corresponding master equation for the reduced atom-cavity density
matrix~\citep{gardiner1985handbook}:
\begin{align}
	\frac{\mathrm{d}}{\mathrm{d}t}\rho=&\frac{1}{\mathrm{i}\hbar}\left[ H, \rho \right]+\mathcal{L}\left[ \rho
	\right]\ ,
\end{align}
with the Liouvillian
\begin{align}
	\mathcal{L}\left[ \rho \right]
	=&\mathcal{L}_{\text{cavity}}\left[ \rho
	\right]+\mathcal{L}_{\text{spont}}\left[ \rho \right]\notag\\
	=&\kappa\left( \bar{n}+1 \right)\left(2a\rho a^{\dagger }- a^{\dagger
	}a\rho-\rho a^{\dagger } a  \right)\notag\\
	&+\kappa\bar{n}\left( 2a^{\dagger }\rho a -a a^{\dagger}\rho-\rho a a^{\dagger }
	\right)\notag\\		
	&-\frac{\gamma_{a}}{2}\sum_{j=1}^{N}\left(\sigma_j^{+}\sigma_j^{-}\rho +\rho
	\sigma_j^{+}\sigma_j^{-}-2\sigma_j^{-}\rho \sigma_j^{+}	\right)\ .
	\label{:master1}
\end{align}
We assumed here that direct thermal excitations of the atoms can be neglected due to the weak
coupling of the hyperfine transition to the environment. The only significant influx of thermal energy
thus occurs via the cavity input-output couplers (mirrors).
Note that the part of the Liouvillian describing spontaneous emission reflects the
assumption that the atoms are coupled to $N$ statistically independent reservoirs. The
main reason for this treatment is that the decay rate $\gamma_{a}$ summarizes the very
small decay rate of atomic excitations and the loss rate of atoms from the trap. Since the
loss of individual atoms from the trap is a noncollective process, the independent
reservoirs assumption is advisable. This part of the Liouvillian cannot be written in
terms of collective operators, and therefore it will not conserve
$J$~\citep{carmichael1999statistical}. Therefore, states with $J<N/2$, including dark
states, become accessible.

%%%%%%%%%%%%%%%%%%%%%%%%%%%%%%%%%%%%%%%%%%%%%%%%%%%%%%%%%%%%%%%%%%%%%%%%%%%%%%%%%%%%%%%%%%%%%%%%%%%%%%%%%%%%%%%%

\section{Signatures of strong coupling}\label{sec:signatures}

A decisive first step toward applications of such system is the precise
characterization and determination of their limits. In particular, the experimental confirmation of
sufficiently strong atom-field coupling compared to the inherent decoherence processes is of vital
importance. This has to be seen in connection with extra limitations induced by thermal
photons in the mode, which in contrast to optical setups play an important role in the
microwave domain.
We thus need reliable methods to determine the atom number, their
effective coupling strength, and noise properties. In particular, we want to find the
minimum temperature requirements that would make it possible to observe strong coupling.

\subsection{Numerical solution for small particle number}\label{sec:num_small}

To get some first qualitative understanding of finite $T$ effects, we study the coupled
atom-field dynamics in the regime of strong coupling under the influence of thermal
photons based on the direct numerical solution of the master equation. Of course, here we have
to resort to the limit of only a few atoms with increased coupling per particle.
Nevertheless, at least the qualitative influence of thermal photons will become visible.
For the practical implementation, we rely on the quantum optics toolbox for Matlab to
explicitly calculate the dynamics of the density matrix~\citep{tan1999computational},
which allows straightforward implementation of the Hamiltonian in Eq.~\eqref{eq:tavis}
formulated in terms of the collective operators.

\begin{figure}[!]
\centering
\subfigure{
\includegraphics[width=0.5\textwidth]{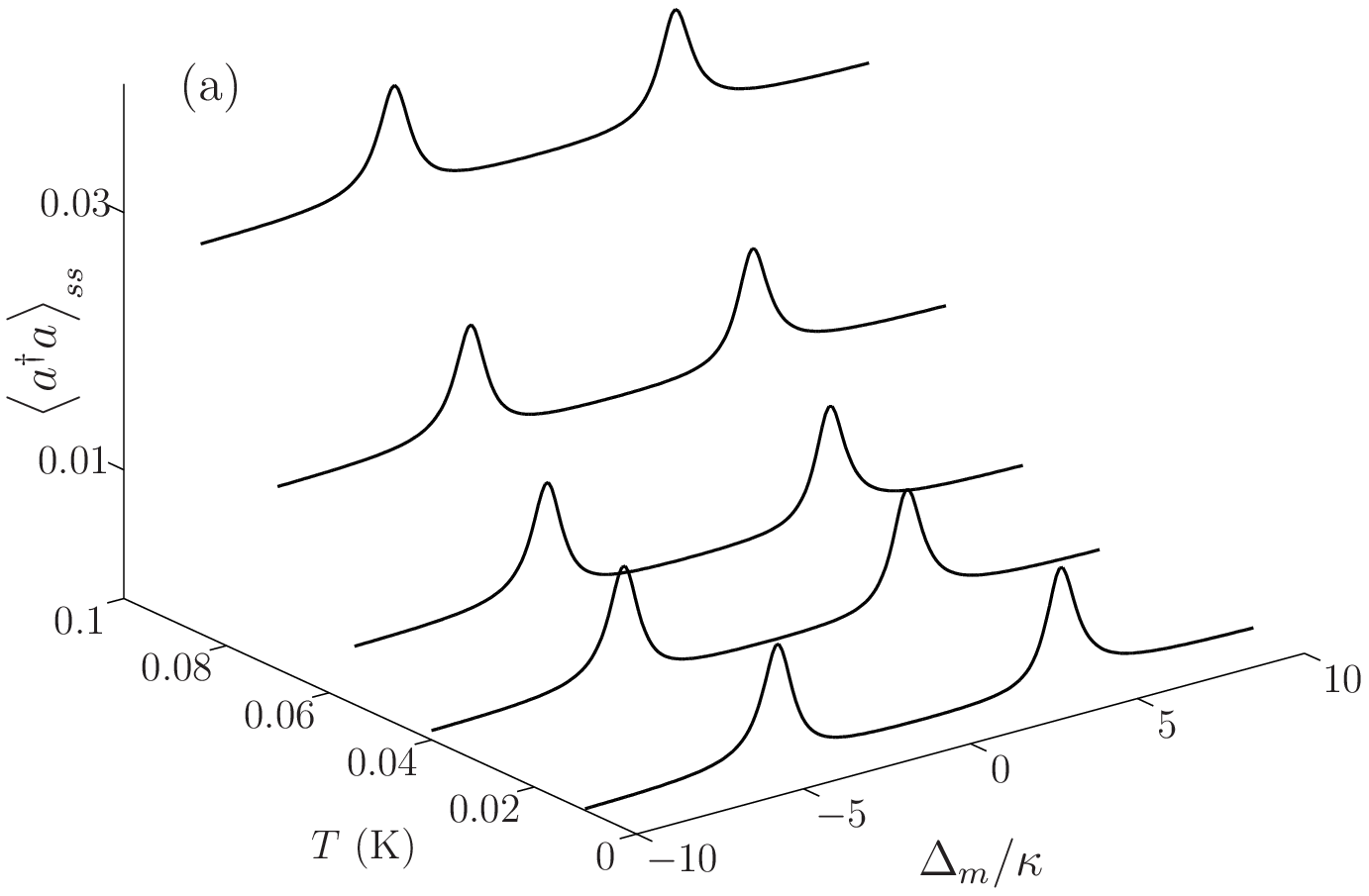}

}
\subfigure{
\includegraphics[width=0.5\textwidth]{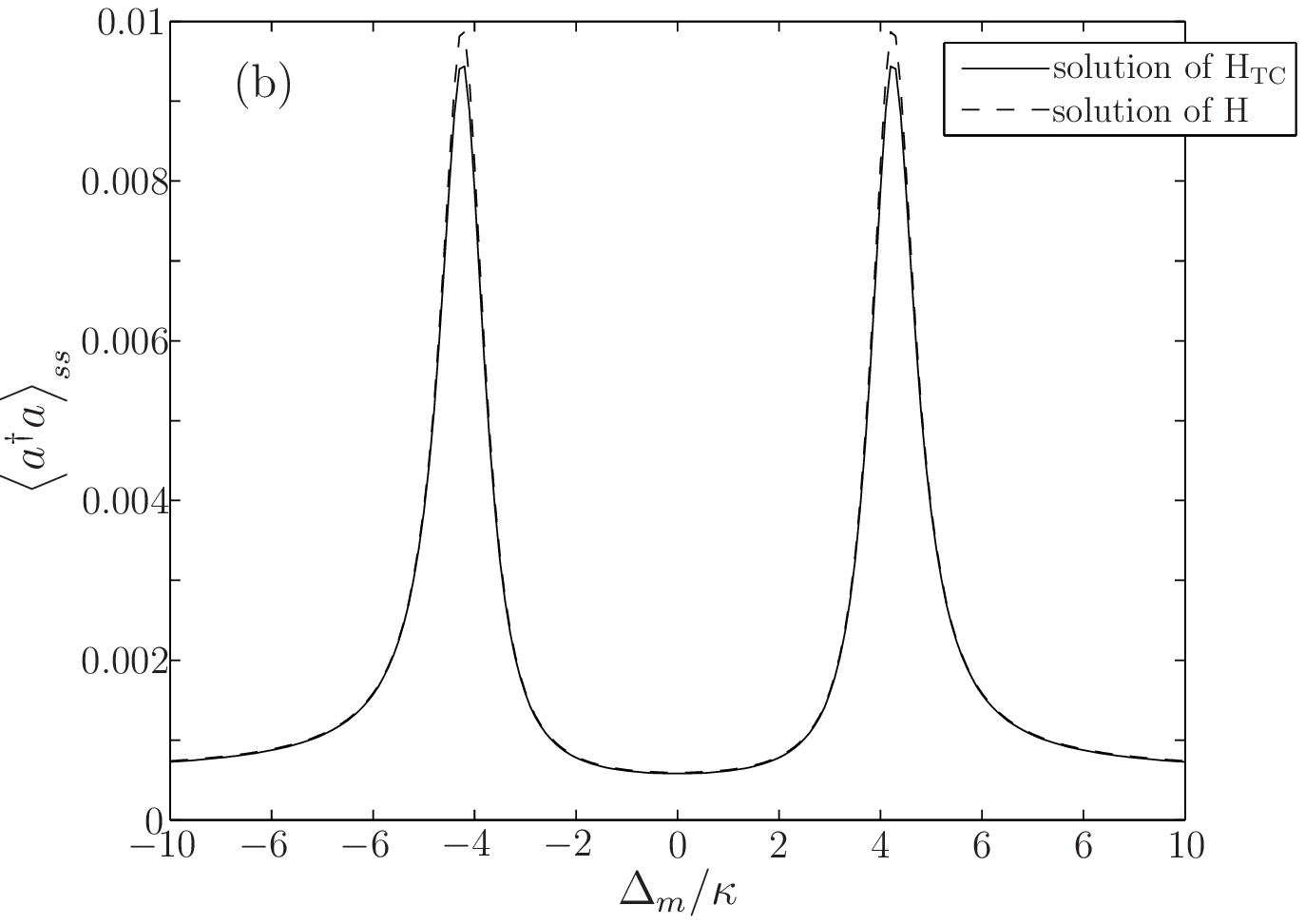}
}
\caption{$\left( \text{a} \right)$ Steady-state number of photons $\left< a^{\dagger }a
\right>_{ss}$ in the pumped cavity for different detunings
$\Delta_{m}=\omega_{m}-\omega_{l}$ and
	different temperatures, obtained from Hamiltonian $\operatorname{H}_{\text{TC}}$.
	The parameters chosen were $\kappa=1$, $N=2$, $g=3$, $\gamma_{a}=0.05$, $\eta=0.1$.
	With increasing temperature the two peaks indicating strong coupling are superimposed by
	thermal photons. To compare the dynamics of $\operatorname{H}_{\text{TC}}$ and
	$\operatorname{H}$, we plot both results for $T=\unit{0.04}{\kelvin}$
	($\bar{n}=3\cdot10^{-4}$) in $\left( \text{b}
	\right)$.}
\label{fig:transspec}
\end{figure}

The cavity is pumped by a coherent microwave field with frequency $\omega_{l}$ and
strength $\eta$, which in
the frame rotating with $\omega_l$ is represented in the Hamiltonian by the additional
term $\operatorname{H}_{p}= \mathrm{i}\hbar \left(\eta a^{\dagger }-\eta^{*}a \right)$.
From the stationary solution, we then determine the steady-state  photon number in the
cavity for different frequencies of the pump field to determine the central system
resonances, where the pump frequency matches the eigenfrequencies $\omega_{m}\pm
g_{\text{eff}}$ of the coupled ensemble-cavity system. At zero temperature and weak
pumping, we get the well-known vacuum Rabi splitting showing two distinct resonances
separated by $2 g_{\text{eff}}$. With increasing temperature and number of thermal
photons, these two peaks will get increasingly broadened and reside on a broad
background.  Figure~\ref{fig:transspec}(a) illustrates the effect.

To compare the dynamics obtained from the restricted Tavis-Cummings Hamiltonian
$\operatorname{H}_{\text{TC}}$ in Eq.~\eqref{eq:tavis} with the dynamics of Hamiltonian
$\operatorname{H}$ in Eq.~\eqref{eq:hamiltonian_full}, we compare the results for both
cases in Fig.~\ref{fig:transspec}(b). Even for the rather small atom numbers chosen here,
the difference and thus the influence of the nonsymmetric states is hardly visible in
this observable.\par
\begin{figure*}[!]
	\begin{center}
		\includegraphics[width=\textwidth]{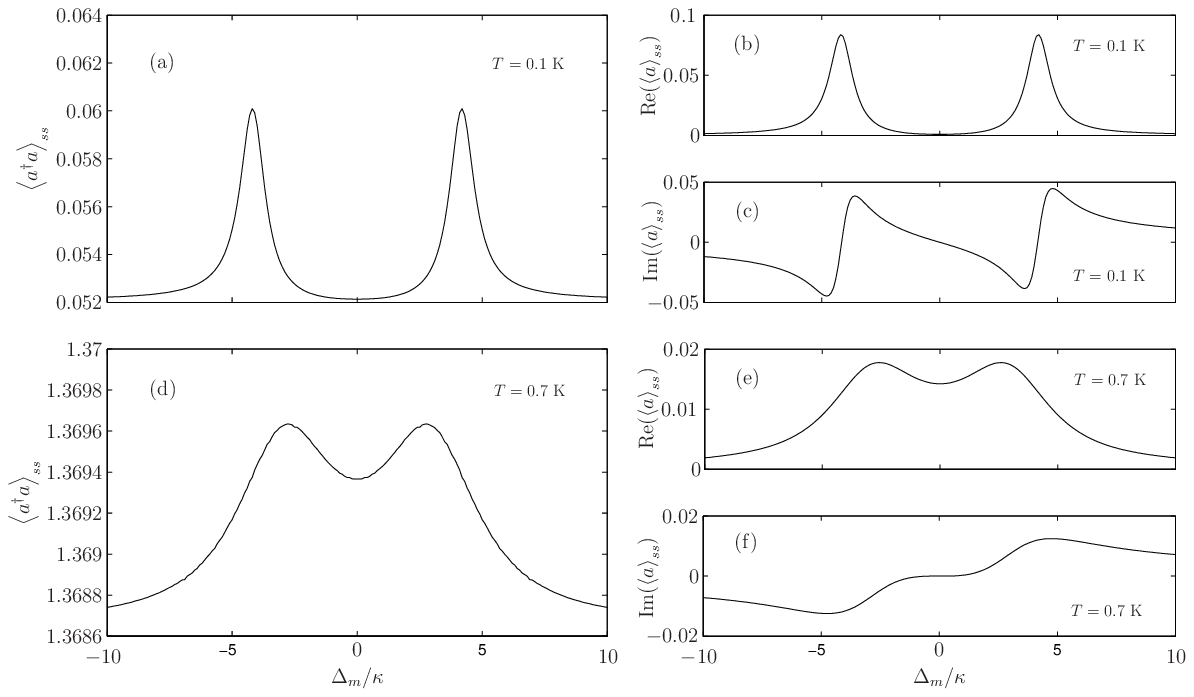}
			\end{center}
	\caption{Effects of higher temperatures: For $T=\unit{0.1}{\kelvin}$,
	 $\left( \text{a} \right)$ shows the steady-state number of photons, whereas
	$\left( \text{b} \right)$ and $\left( \text{c} \right)$ show the real and imaginary part
	of the steady state field in the cavity. The peaks in the photon number start to broaden
	and finally vanish. The amplitude of the field shows similar behavior. Panels $\left(
	\text{d} \right)$, $\left( \text{e} \right)$, and $\left( \text{f} \right)$ show the same
	quantities for $T=\unit{0.7}{\kelvin}$. The remaining parameters were chosen as in
	Fig.~\ref{fig:transspec}.}
		\label{fig:uebersicht_transspec_g_3_Te_01_03_1_N6_Na2_wl_90_01_110}
\end{figure*}

 The peaks in the photon number in principle stay visible also for higher
temperatures, but they start to broaden and finally vanish. Regardless, the detection on the
thermal background gets technically more challenging. The intracavity steady-state
amplitude of the field shows a similar behavior and a determination of the splitting
becomes increasingly impossible, despite the fact that phase-sensitive detection (homodyne)
can help. An example is shown in
Fig.~\ref{fig:uebersicht_transspec_g_3_Te_01_03_1_N6_Na2_wl_90_01_110}. This effect is
expected to be less important if the ensemble contains a large number of atoms. However,
this regime is not accessible for direct numerical simulations and we have to develop
alternative semianalytic approaches.

%\clearpage

%%%%%%%%%%%%%%%%%%%%%%%%%%%%%%%%%%%%%%%%%%%%%%%%%%%%%%%%%%%%%%%%%%%%%%%%%%%%%%%%%%%%%%%%%%%%%%%%%%%%%%%%%%%%%%%

\subsection{Truncated cumulant expansion of collective
observables}\label{sec:cumulantexpansion}

To overcome the system size restrictions of a direct numerical solution, we now turn to an
alternative approach that does not rely on the simulation of the dynamics of the whole density
matrix. Instead, we derive a system of coupled differential equations for the expectation
values of the relevant system variables.  The inversion of atom $i$ then obeys
\begin{align}
	\frac{\mathrm{d}}{\mathrm{d}t}\left< \sigma_i^{z}
	\right>&=\operatorname{Tr}\{\sigma_i^{z}\frac{\mathrm{d}}{\mathrm{d}t}\rho\}\notag\\
	&=-\mathrm{i}2g\left( \left< \sigma_i^{+}a
	\right>-\left< \sigma_i^{-}a^{\dagger } \right> \right)-\gamma_{a}\left( \left<
	\sigma_i^{z}
	\right> +1 \right)
\end{align}
which couples to $\left< \sigma_i^{+}a \right>$ and $\left< \sigma_i^{-}a^{\dagger }
\right>=\left< \sigma_i^{+}a \right>^{*}$. We assume that all atoms are equal, which allows
us to replace $\left<\sigma_i^{z}\right>$ with $\left<\sigma_1^{z}\right>$. Expectation values for pairs
of different atoms like $\left< \sigma_{i}^{+}\sigma_{j}^{-} \right>$ can be replaced with
$\left< \sigma_{1}^{+}\sigma_{2}^{-} \right>$.

While these equations are exact in principle, the procedure ultimately leads to an
infinite set of coupled equations. We thus have to start approximations and truncate this
set at a chosen point, neglecting higher-order
cumulants~\citep{kubo1962generalized,meiser2009pml}. The truncation has to be carefully
chosen and tested in general. Here we stop at third order, which in similar
situations has proven to be well suited to describe the essential
correlations~\citep{meiser2009pml}.  \par The
expansion for an expectation value of the form $\left< ab \right>$ is the well-known
relation $\left< ab \right>=\left< ab \right>_{c}+\left< a \right>\left< b \right>$, with
$\left< ab \right>_{c}$ being the covariance between $a$ and $b$. Along this line, one
expands third-order terms in the form
\begin{eqnarray}
	\left< abc \right>&=&\underbrace{\left< abc \right>_{c}}_{\text{neglected}}+\left< ab
	\right>_{c}\left< c \right>+\left< ac
	\right>_{c}\left< b \right>\notag\\&&+\left< bc \right>_{c}\left< a \right>+\left< a \right>\left< b
	\right>\left< c \right>\ .
	\label{eq:expand3}
\end{eqnarray}
We make one exception in this expansion when it comes to the quantity $\left< a^{\dagger }a
\sigma_{1}^{z} \right>$; the reason for this is discussed in Appendix~\ref{app:validity}.

\begin{figure*}[!]
	\begin{center}
		\includegraphics[width=\textwidth]{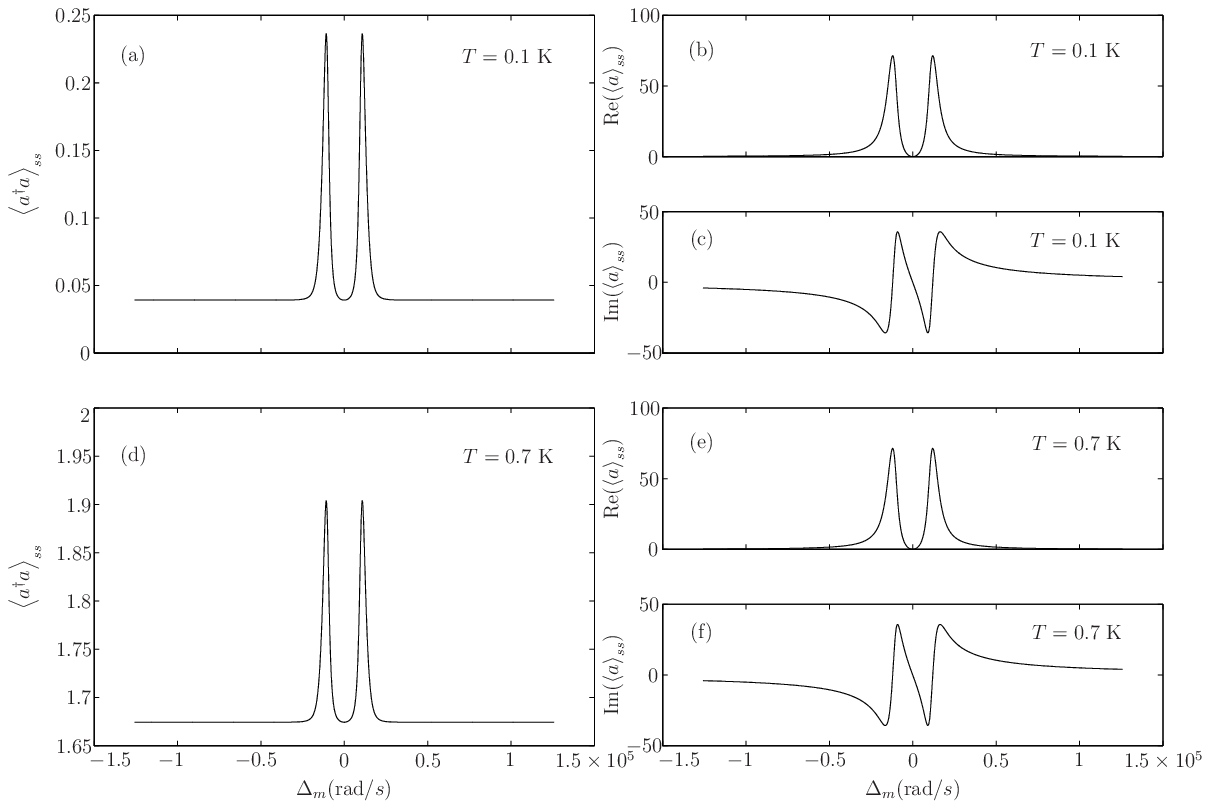}
					\end{center}
	\caption{The steady-state field and photon number in the cavity for different
	frequencies of the pump. The size of the ensemble was chosen to be $N=10^{5}$ and we set
	$\kappa=7\cdot10^{3}$, $\eta=5\cdot10^{5}$, $\gamma_{a}=0.3$ and $g=40$. Panel $\left(
	\text{a} \right)$ shows the steady-state number of photons, whereas $\left( \text{b}
	\right)$ and $\left( \text{c} \right)$ show the real and imaginary parts of the
	steady-state field in the cavity for $T=\unit{0.1}{\kelvin}$, which corresponds to
	$\bar{n}=0.04$. The remaining figures show the same quantities for
	$T=\unit{0.7}{\kelvin}$, corresponding to $\bar{n}=1.67$. The results show that for a
	sufficiently large number of atoms strong coupling remains observable despite of the
	presence of thermal photons.}
	\label{fig:uebersicht_longtime_photons_stst_do_radiate_third_N10to5_kappa7000_g40_T01_07_eta_5_10to5}
\end{figure*}

The number of equations depends on the order of the cumulants we wish to keep track of.
Furthermore, the problem is greatly simplified if there is no coherent input field driving our
cavity. In this case, no defined phase exists in our system, so we can assume that
$\left< a \right>=\left< a^{\dagger } \right>=\left< \sigma^{\pm}_1 \right>=0
$. Note that while for a single system trajectory a coherent field can build up as in a laser, for an
average over many realizations the preceding assumption holds. For a covariance like $\left<
\sigma_1^{+}a \right>_{c}=\left<
\sigma_1^{+}a \right>-\left< \sigma^{+}_{1}\right>\left< a
\right>$, we therefore find $\left< \sigma_1^{+}a \right>=\left<
\sigma_1^{+}a \right>_{c}$. The four remaining equations are
\begin{eqnarray}
	\frac{\mathrm{d}}{\mathrm{d}t}\left< \sigma_1^{z} \right>&=&-\mathrm{i}2g\left( \left< \sigma_1^{+}a
	\right>-\left< \sigma_1^{-}a^{\dagger } \right> \right)\notag\\&&-\gamma_{a}\left( \left<
	\sigma_1^{z}
	\right> +1 \right)\label{eq:sz}
\end{eqnarray}

\begin{eqnarray}
	\frac{\mathrm{d}}{\mathrm{d}t}\left< a\sigma_1^{+} \right>&=&-\left(
	\kappa+\frac{\gamma_{a}}{2}+\mathrm{i}\left(\omega_{m}-\omega_a
	\right)\right)\left< a \sigma_1^{+} \right>\nonumber\\
	&&-\mathrm{i}g\biggl( \frac{\left< \sigma_1^{z} \right> +1}{2}+\left< a^{\dagger }a
	\right>\left< \sigma_1^{z}
	\right>\biggr.\notag\\&&\qquad \biggl.+\left( N-1 \right)\left<
	\sigma_1^{+}\sigma_2^{-}
	\right>\biggr)\label{eq:asp}
	\end{eqnarray}

\begin{eqnarray}
	\frac{\mathrm{d}}{\mathrm{d}t}\left< a^{\dagger }a \right>&=&-\mathrm{i}gN\left( \left< a^{\dagger
	}\sigma_1^{-} \right>-\left< a \sigma_1^{+} \right> \right)\notag\\
	&&-2\kappa\left< a^{\dagger}a \right>+2\kappa\bar{n}\label{eq:ada}
\end{eqnarray}

\begin{eqnarray}
	\frac{\mathrm{d}}{\mathrm{d}t}\left< \sigma_1^{+}\sigma_2^{-} \right>&=& -\gamma_{a}\left< \sigma_1^{+}\sigma_2^{-}
	\right>\notag\\&&-\mathrm{i}g\left<
	\sigma_1^{z}
	\right>\left( \left<
	\sigma_1^{-}a^{\dagger }
	\right>-\left< \sigma_1^{+}a \right> \right)\label{eq:spsm1}\ .
\end{eqnarray}
To inject energy into our system without losing the property of having no defined phase, we
can introduce an incoherent pump of the atoms. In essence, this gives an additional term in
the Liouvillian very much resembling spontaneous emission in opposite direction.
Formally, it reads $-\frac{w}{2}\sum_{j=1}^{N}\left(\sigma_j^{-}\sigma_j^{+}\rho +\rho
\sigma_j^{-}\sigma_j^{+}-2\sigma_j^{+}\rho \sigma_j^{-}	\right)$, where $w$ denotes the
rate of the pump. The modifications of Eqs.~\eqref{eq:sz}--\eqref{eq:spsm1} narrow down to
the replacement of $\gamma_{a}$ with $\gamma_{a}+w$ and of $-\gamma_{a}\left( \left< \sigma_1^{z}\right>
+1 \right)$ with $-\left( \gamma_{a}+w \right)\left( \left< \sigma_1^{z} \right> +
\frac{w-\gamma_{a}}{w+\gamma_{a}}\right)$ in Eq.~\eqref{eq:sz}.\par

Introducing a coherent pump leads to a larger set of 13 coupled equations for the
quantities $\left< a \right>$, $\left< \sigma_{1}^{z}\right>$, $\left< \sigma_{1}^{+}
\right>$, $\left< a \sigma_{1}^{+} \right>_{c}$, $\left< a \sigma_{1}^{z} \right>_{c}$,
$\left< \sigma_{1}^{+}\sigma_{2}^{-} \right>_{c}$, $\left< a^{\dagger }a \right>_{c}$,
$\left< a\sigma_{1}^{-} \right>_{c}$, $\left< a^{\dagger }a^{\dagger } \right>_{c}$,
$\left< \sigma_{1}^{-}\sigma_{2}^{-} \right>_{c}$, $\left< \sigma_{1}^{z}\sigma_{2}^{+}
\right>_{c}$, $\left< \sigma_{1}^{z}\sigma_{2}^{z} \right>_{c}$ and $\left< a^{\dagger }a
\sigma_{1}^{z} \right>$; for details, see Appendix~\ref{app:A}. In this case we transform into a
rotating frame with respect to the frequency of the pump $\omega_{l}$. This results in
$\Delta_{m}=\omega_{m}-\omega_{l}$ for the detuning of the cavity and
$\Delta_{a}=\omega_{a}-\omega_{l}$ for the detuning of the atoms with respect to the pump
frequency.

In general, the set of equations is too complex for a direct analytical solution and has to
be  integrated numerically. In this way we obtain the steady-state expectation values of
relevant observables as the occupation number of the cavity $\left< a^{\dagger }a\right>$
or the inversion of the ensemble $\left< \sigma_{1}^{z} \right>$.

\begin{figure}[h]
	\begin{center}
		\includegraphics[width=0.4\textwidth]{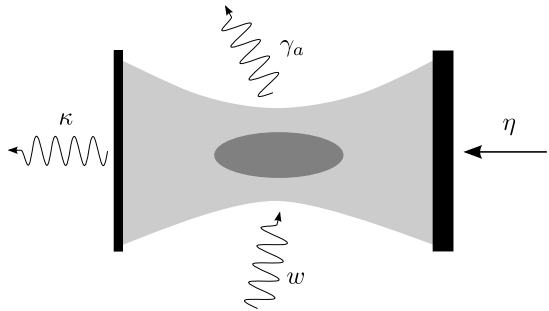}
	\end{center}
	\caption{To simplify matters, we depict the cavity as a Fabry-Perot cavity which can be
	pumped through a mirror with high reflectivity. The observation of the dynamics is carried
	out using the second mirror, which has a lower reflectivity. Additionally, we can pump the
	ensemble incoherently from the side.}
	\label{fig:cavity}
\end{figure}

To compare the results of the obtained equations with the results in
Sec.~\ref{sec:num_small}, we plot the steady-state number of photons and the field in the
cavity for different frequencies of the pump laser in
Fig.~\ref{fig:uebersicht_longtime_photons_stst_do_radiate_third_N10to5_kappa7000_g40_T01_07_eta_5_10to5}.
We clearly see that effective strong coupling appears for a sufficiently large number of
weakly coupled atoms and can stay visible also at higher temperatures. Further discussion
is given in Sec.~\ref{sec:driven_mode}. In Fig.~\ref{fig:cavity} we schematically depict
the setup including the described loss and pump processes.

%\clearpage
\subsection{Cavity output spectrum at finite temperature}\label{sec:Spectrum}

Naturally, the total field intensity in the cavity is only part of the story and significant physical
information can be obtained from a spectral analysis of the transmitted field. Using the
quantum regression theorem, the spectrum of the light transmitted through one of the
mirrors can be expressed in terms of the Fourier transform of the corresponding
autocorrelation function of the field amplitude.

At finite $T$ this is not the full story, and to obtain the actual spectrum of the light
impinging on the detector, one has to include the thermal photons in the output mode
reflected from the cavity. Hence, the normalized first-order correlation function of the
field outside the cavity will have additional contributions from the correlation function
of the thermal field, as well as of the correlation between the thermal field and the
cavity field~\citep{carmichael1987spectrum}. The latter is causing interference effects
between cavity field and the thermal field. The correlations between reservoir operators
and cavity operators can be expressed in terms of averages involving cavity operators
alone~\citep{carmichael1999statistical,carmichael1987spectrum}.  For a cavity radiating
into a thermal reservoir, we find for the normalized first-order correlation function
\begin{eqnarray}
	g\left( \tau \right)&=&\frac{1}{\mathcal{N}}\Biggl\{ \frac{1}{2\pi g\left( \omega
	\right)}\left< r_{f}^{\dagger}\left( 0 \right)r_{f}\left( \tau \right)
	\right>\Biggr.\notag\\
	&&\Biggl.+2\kappa\left[ \lim_{t\rightarrow \infty}\left< a^{\dagger}\left( t
	\right)a\left( t+\tau \right) \right> \right]\Biggr.\notag\\
	&&\Biggl. +2\kappa\bar{n}\left(
	\omega_{m},T
	\right)\left[ \lim_{t\rightarrow \infty}\left< \left[
	a^{\dagger}\left( t \right),a\left( t+\tau \right) \right] \right> \right]\Biggr\}\ ,\quad
	\label{eq:g}
\end{eqnarray}
with
\begin{align}
	\mathcal{N}=\frac{1}{2\pi g\left( \omega \right)}\left< r_{f}^{\dagger}r_{f}
	\right>+2\kappa\left( \left< a^{\dagger}a \right>_{\text{ss}}-\bar{n}\left(
	\omega_{m},T
	\right) \right)\ .
	\label{eq:norm}
\end{align}
Here, $r_{f}$ denotes the annihilation operator of a reservoir photon and $g\left( \omega
\right)$ denotes the density of states in the reservoir. Equations for the correlation
functions in Eq.~\eqref{eq:g} can be obtained via the quantum regression theorem. The
resulting system of coupled equations is Laplace transformed to give the contributions to
the spectrum that arise from the reservoir, the cavity and cavity-reservoir interference.
The initial conditions necessary for the Laplace transform are the steady-state values
obtained either numerically for the coherently pumped cavity or analytically (see
Sec.~\ref{sec:laser}).\par
\begin{figure*}[!]
	\begin{center}
		\includegraphics[width=0.7\textwidth]{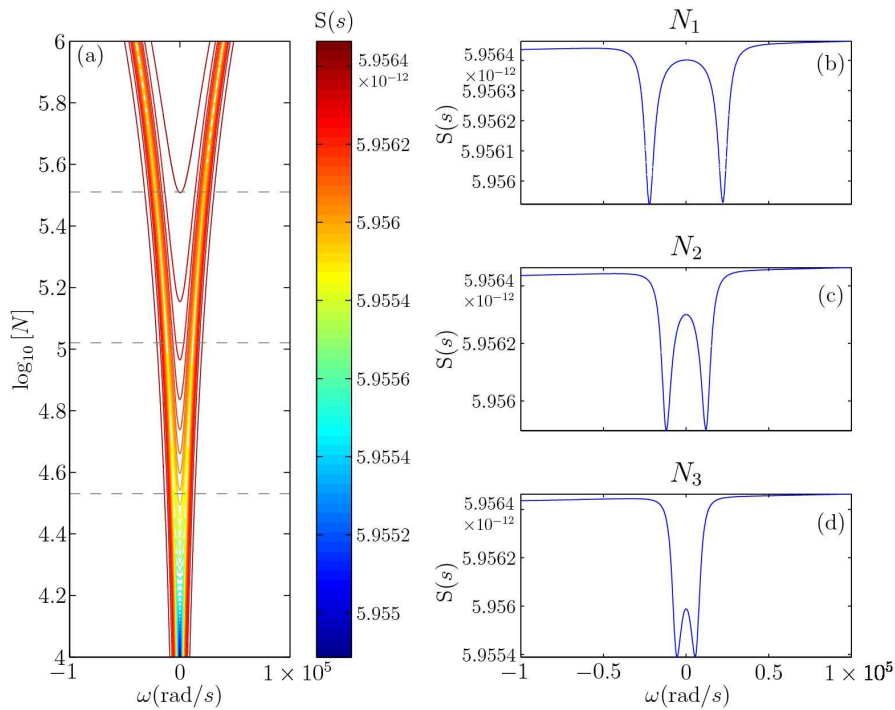}
			\end{center}
			\caption{(Color online) Overview of the transmitted spectrum $\operatorname{S}$ for
			different sizes of the ensemble (a). The temperature of the cavity is set to
			$T=\unit{0.1}{\kelvin}$ ($\bar{n}=0.04$). The remaining parameters were chosen to be
			$\kappa=7\cdot 10^3$, $\gamma_{a}=0.3$, $g=40$, $\omega_{a}=\omega_{m}=2\pi \cdot
			\unit{6.83}{\giga \hertz}$. The dips in the spectrum indicate that thermal photons
			are absorbed by the ensemble and re-emitted into modes other than the cavity mode.
			The increasing distance between the absorption dips reflects the increasing number
			of atoms. In panels (b)-(d) we depict the spectra at $N_1=3.2 \cdot 10^{6}$, $N_2=1
			\cdot 10^{6}$, and $N_3=3.4\cdot 10^{5}$, indicated in (a) by the dashed horizontal
			lines.}
		\label{fig:S_trans_thermal_N10to4_10to6_T01_kappa7000_g40_gamma03_w0}
\end{figure*}

We show the spectrum of the cavity without any pump, coherent or incoherent, but
with $T=\unit{0.1}{\kelvin}$ in
Fig.~\ref{fig:S_trans_thermal_N10to4_10to6_T01_kappa7000_g40_gamma03_w0}. The spectrum
shows absorption dips at the frequencies of the coupled ensemble-cavity system.  Some
thermal photons that leak into the cavity are absorbed and lost into modes other than the
cavity mode. In this form the thermal field is a broadband probe of resonant system
absorption.

\subsection{Cooling the field mode with the atomic ensemble }\label{sec:cooling}

\begin{figure}[!]
	\begin{center}
		\includegraphics[width=0.5\textwidth]{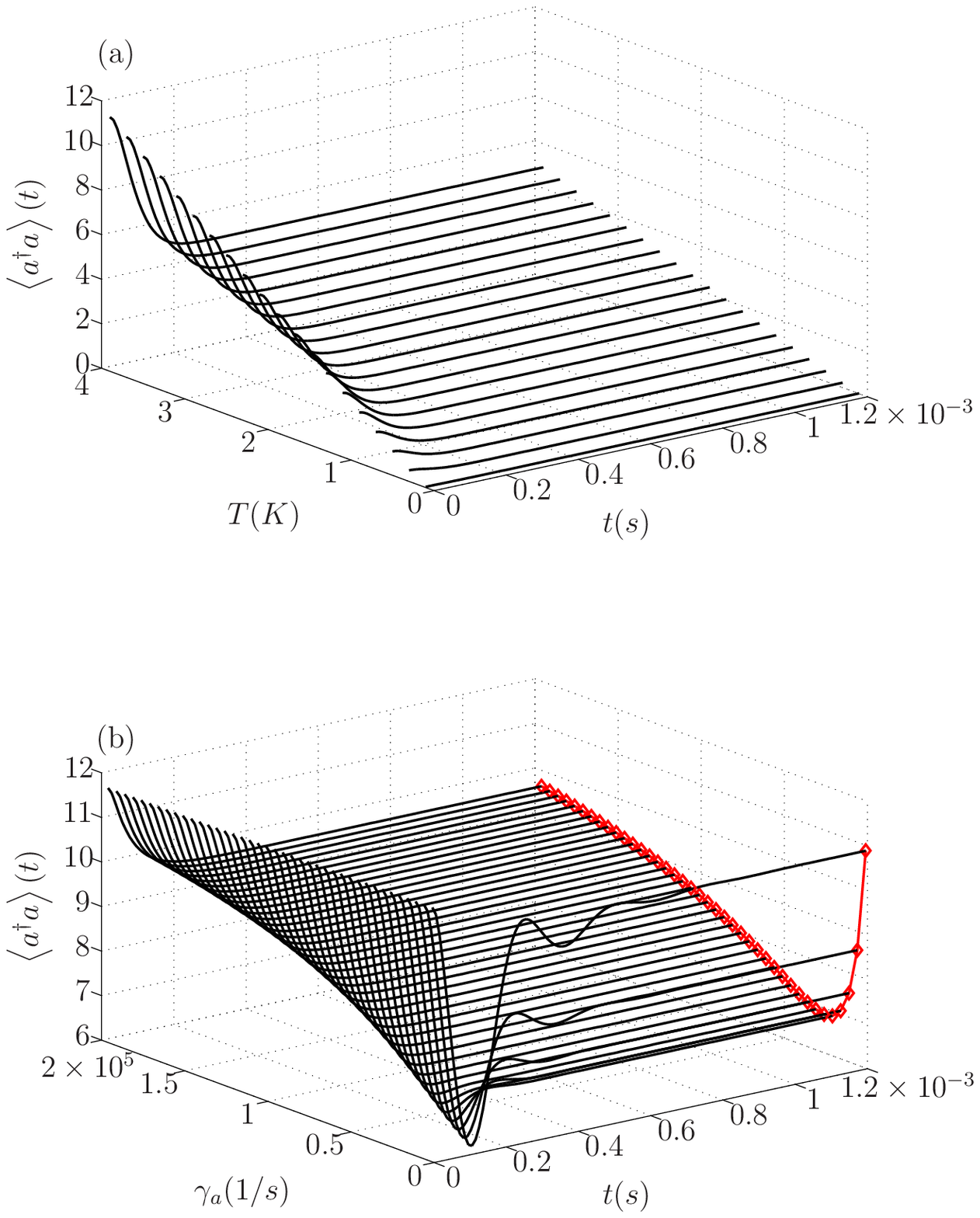}
			\end{center}
	\caption{(Color online) Loss of photons from the cavity mode. $\left( \text{a} \right)$ Dynamics of the
	occupation of the mode for different temperatures and $\gamma_{a}=5\cdot10^4$. A constant
fraction of the photons is removed from the cavity mode. $\left( \text{b} \right)$ For
fixed $T=\unit{4}{\kelvin}$ ($\bar{n}=11.7$), the loss rate of the atoms is varied between $\gamma_{a}=1\cdot 10^{3}$
and $2\cdot 10^5$. The steady-state number of photons shows that there is an optimal
loss rate. The red curve (with diamond markers) corresponds to the expected number of photons according to
Eq.~\eqref{eq:est_ada}. The remaining parameters were chosen to be $\kappa=7\cdot	10^3$, $g=40$,
$\omega_{a}=\omega_{m}=2\pi \cdot\unit{6.83}{\giga\hertz}$, $N=10^5$.}
	\label{fig:cool_cavity_vary_gamma_and_T}
\end{figure}

The spectra depicted in
Fig.~\ref{fig:S_trans_thermal_N10to4_10to6_T01_kappa7000_g40_gamma03_w0} show a weak loss
of thermal photons from the coupled ensemble-cavity system. Cavity photons are absorbed
and sometimes scattered into a mode other than the cavity mode. As the ensemble can be
nearly perfectly optically pumped into a particular state, its effective temperature is
close to zero and hence well below the mode temperature. A relative purity of the ensemble
of $10^{-5}$ corresponds to $T\sim \unit{28}{\milli\kelvin}$, where we used $\hbar
\omega_{a}/k_{b}T=\operatorname{ln}\left( 10^{-5} \right)$ with $\omega_{a}/ 2 \pi
=\unit{6.83}{\giga\hertz}$. This leads to the question as to what extent the thermal
occupation of the mode can be reduced by thermal contact between the two systems via
such energy transfer and loss. In Fig.~\ref{fig:cool_cavity_vary_gamma_and_T}$\left(
\text{a} \right)$ we show the dynamics of the photon number in the mode at different
temperatures after putting the systems into contact. In
Fig.~\ref{fig:cool_cavity_vary_gamma_and_T}$\left( \text{b} \right)$ we
consider different loss or decay rates $\gamma_{a}$ of the excited atoms. In practice one
could think of coupling to an magnetically untrapped atomic state or adding some repumping
mechanism to increase this intrinsically very low rate. The dynamics is found numerically
by integrating Eqs.~\eqref{eq:sz}--\eqref{eq:spsm1}. To see the effect for increasing
temperature in Fig.~\ref{fig:cool_cavity_vary_gamma_and_T}$\left( \text{a} \right)$, we initialize the ensemble with all atoms in
the ground state, whereas the mode contains $\bar{n}\left(\omega_{m}, T \right)$ photons.
The decay rate of the atoms is chosen to be $\gamma_{a}=5\cdot10^4$. With increasing
temperature the initial number of photons also increases. Due to coherent transfer and
decay via the atoms, a constant fraction of the photons is removed from the cavity mode. In
Fig.~\ref{fig:cool_cavity_vary_gamma_and_T} $\left( \text{b} \right)$ we show the same
effect except that we now keep the temperature fixed to $T=\unit{4}{\kelvin}$ and vary the
decay of the atoms $\gamma_{a}=1\cdot 10^{3}\dots 2\cdot 10^5$. The steady state of the
photon number strongly depends on $\gamma_{a}$. The red curve (with diamond markers) is the expected number of
photons remaining in the cavity
\begin{align}
	\left< a^{\dagger }a \right>_{ss}=\bar{n}\left(\omega_{m}, T
	\right)-\frac{1}{2\kappa}N\gamma_{a}\left( \frac{1+\left< \sigma_{1}^{z}
	\right>_{\text{ss}}}{2} \right)
	\label{eq:est_ada}
\end{align}
which coincides with the numerical results. The inversion $\left< \sigma_{1}^{z}
	\right>_{\text{ss}}$ can be calculated analytically from Eqs.~\eqref{eq:sz}--\eqref{eq:spsm1}. The loss of thermal photons is proportional to
the loss rate $\gamma_{a}$ and the number of atoms in the excited state $N\left(
\frac{1+\left< \sigma_{1}^{z}\right>_{\text{ss}}}{2} \right)$. The latter becomes very
small if $\gamma_{a}$ becomes large. Hence there is an optimal loss rate for each set of
parameters.

%\clearpage

\begin{figure}[t]
	\begin{center}
		\includegraphics[width=0.5\textwidth]{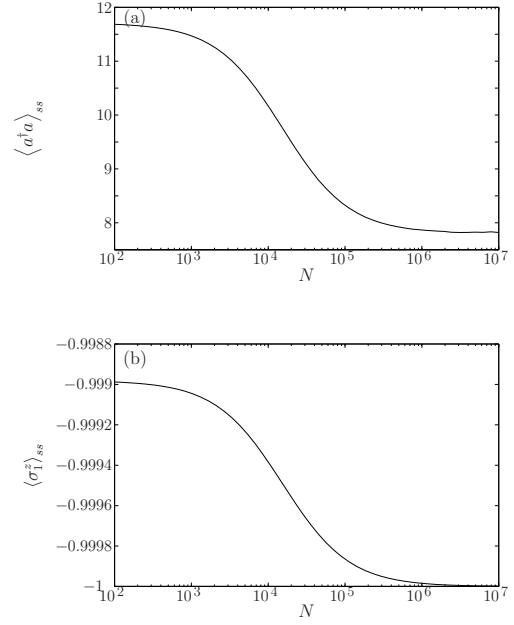}
	\end{center}
	\caption{Steady state of the photon number (a) and the inversion (b) for a loss rate
	$\gamma_{a}=\kappa=7\cdot 10^{3}$. Increasing the number of atoms $N$ leads to a more
	effective removal of thermal photons. The remaining parameters are chosen to be $g=40$,
	$\omega_{a}=\omega_{m}=2\pi \cdot \unit{6.83}{\giga \hertz}$, $T=\unit{4}{\kelvin}$
	($\bar{n}=11.7$).}
	\label{fig:cool_cavity_paper_N10to2_10to7_g40_w0_gamma7000_kappa7000_T4}
\end{figure}

The removal of thermal photons becomes more effective if the number of atoms is increased.
Meanwhile, the inversion of the ensemble also drops since the fraction of excited atoms is
decreased. Overall, the effect is clearly visible (see
Fig.~\ref{fig:cool_cavity_paper_N10to2_10to7_g40_w0_gamma7000_kappa7000_T4}), but it seems that for the actual
parameters here its practical value remains limited. However, with a larger atom number
and more tailored decay rates the method could be employed to reset a particular mode
shortly before starting any quantum gate operation. Note that this treatment of the
cooling process is limited to short time scales since the permanent loss of excitations
via $\gamma_{a}$ involves the loss of atoms from the ensemble. The number of lost atoms
after the time $t$,
approximated by the number of thermal photons that entered the cavity
$t\kappa\bar{n}$, has to be much smaller than the ensemble size $N$, which restricts the
time $t$.

%%? Stimmt dieses Ergebnis mit Gl 22 zusammn (Plot von 10^3 weg)

%\clearpage

\begin{figure*}[t]
	\begin{center}
		\includegraphics[width=0.7\textwidth]{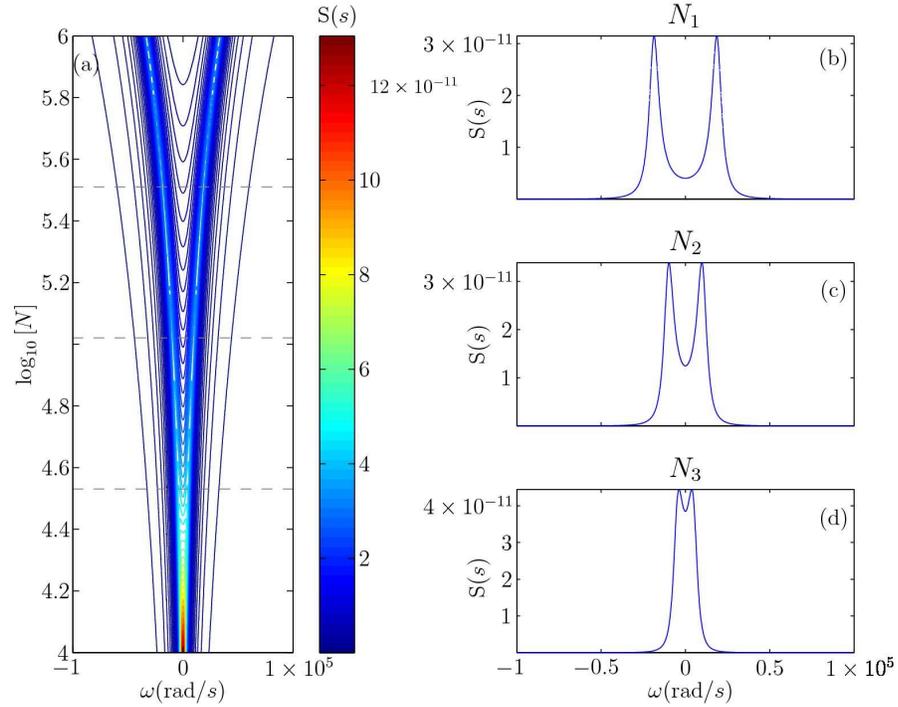}
		\end{center}
	\caption{(Color online) The temperature of the cavity is now set to
	$T=\unit{0.001}{\kelvin}$ ($\bar{n}=0$) and an incoherent pump of the atoms with
	$w=0.05$ is switched on. The spectrum now shows increased emission at the frequencies of
	the coupled system. Again figures (b)-(d) depict the spectra at $N_1=3.2 \cdot 10^{6}$,
	$N_2=1 \cdot 10^{6}$, and $N_3=3.4\cdot 10^{5}$, indicated in (a) by the dashed
	horizontal lines.}
		\label{fig:S_trans_thermal_N10to4_10to6_T0001_kappa7000_g40_gamma03_w005}
\end{figure*}

In the situation where the mode is at $T\approx \unit{0}{\kelvin}$ and the atoms are
subject to incoherent pumping we find increased transmission for the resonance
frequencies (see
Fig.~\ref{fig:S_trans_thermal_N10to4_10to6_T0001_kappa7000_g40_gamma03_w005}). We again
recover the $\sqrt{N}$ dependence of the splitting of the peaks.

\clearpage

\subsection{Coherently driven cavity mode}\label{sec:driven_mode}

An experimentally readily accessible quantity is the cavity field amplitude, which can be
deduced by phase-sensitive (homodyne) detection of the output. This quantity is much less
obscured by random thermal field fluctuations than the spectral intensity in total.  As a
phase reference, we therefore now introduce a coherent phase stable pump of the cavity,
which is again represented in the Hamiltonian by the additional term
$\operatorname{H}_{p}= \mathrm{i}\hbar \left(\eta a^{\dagger }-\eta^{*}a \right)$.

\begin{figure}[!]
	\begin{center}
		\includegraphics{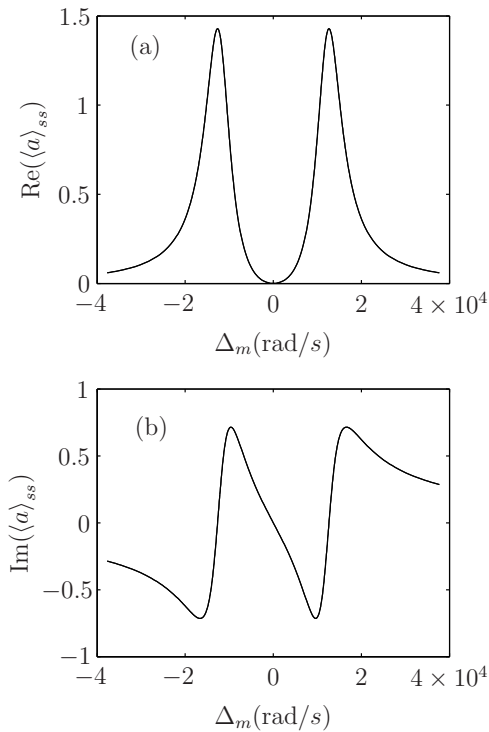}
	\end{center}
	\caption{Steady state field in the driven cavity, real part (a) and imaginary
	part (b). The size of the ensemble is chosen to be $N=10^{5}$. The lines for
	$T=\unit{0.01}{\kelvin}$ ($\bar{n}=0$) and $T=\unit{10}{\kelvin}$ ($\bar{n}=30$) coincide. }
	\label{fig:metaspektrum_gamma03_N10to5_kappa7000_g40_T001_10_eta10to4_Delta_angepasst}
\end{figure}

As mentioned previously, a coherent pump strongly increases the number of nonvanishing
cumulants and at our level of truncation leads to a set of 13 coupled equations, which can
be found in Appendix~\ref{app:A}. Based on this set, we can calculate the stationary real- and
imaginary part of the field in the cavity after transient dynamics. The amplitude of the
field inside the cavity becomes maximal if the frequency $\omega_l$ of the driving laser
hits one of the resonances of the coupled system. As we give a phase reference now the
effect of a higher temperature on the field in the cavity is barely visible, in
particular if we chose a large ensemble of $N=10^5$ atoms (see
Fig.~\ref{fig:metaspektrum_gamma03_N10to5_kappa7000_g40_T001_10_eta10to4_Delta_angepasst}).

\begin{figure}[t]
	\begin{center}
		\includegraphics{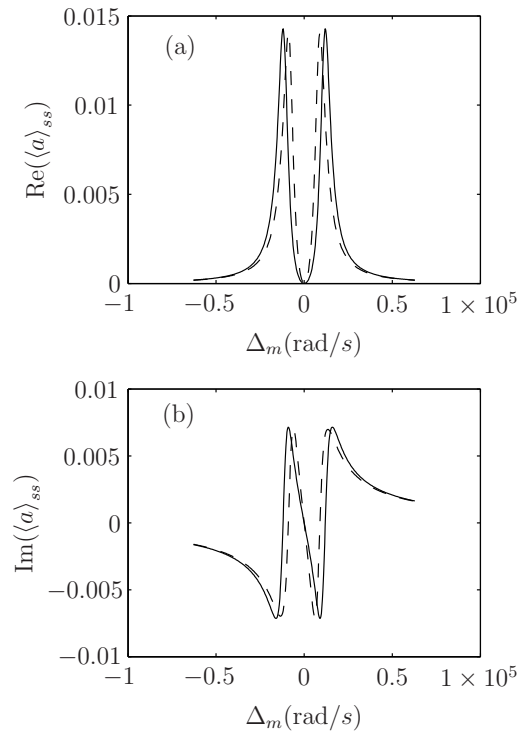}
	\end{center}
	\caption{Steady state field in the driven cavity, real part (a) and imaginary
	part (b), for $T=\unit{0.01}{\kelvin}$
	($\bar{n}=0$) (solid lines) and
	$T=\unit{10}{\kelvin}$ ($\bar{n}=30$) (dashed lines). For the small ensemble with $N=10^2$
	atoms we recover the effects of the thermal photons. To
	compensate for the lower number of atoms, the coupling is chosen to $g=1200$. Otherwise,
	the splitting would be covered by the cavity linewidth. }
	\label{fig:metaspektrum_gamma03_N10to2_kappa7000_g1200_T001_10_eta10to2_Delta_angepasst}
\end{figure}

Note that although not giving the vacuum Rabi splitting, the average atom-field coupling can be still deduced
from these resonances as $g$ enters in their frequency. 
If we go back to a rather small ensemble of $N=10^2$ atoms, the influence of the temperature
becomes visible. To ensure that we still can observe well split levels, which are not
covered by the linewidth of the cavity, we increase the coupling constant $g$ in our
simulation. The results in Fig.~\ref{fig:metaspektrum_gamma03_N10to2_kappa7000_g1200_T001_10_eta10to2_Delta_angepasst} show that thermal effects become visible in the field only if the
number of thermal photons is not negligible compared to $N$.

%\clearpage

%%%%%%%%%%%%%%%%%%%%%%%%%%%%%%%%%%%%%%%%%%%%%%%%%%%%%%%%%%%%%%%%%%%%%%%%%%%%%%%%%%%%%%%%%%%%%%%%%%%%%%%%%%%%%

\subsection{Spectrum of the coherently driven cavity}\label{sec:spec_driven}

	\begin{figure*}[!]
		\begin{center}
			\includegraphics{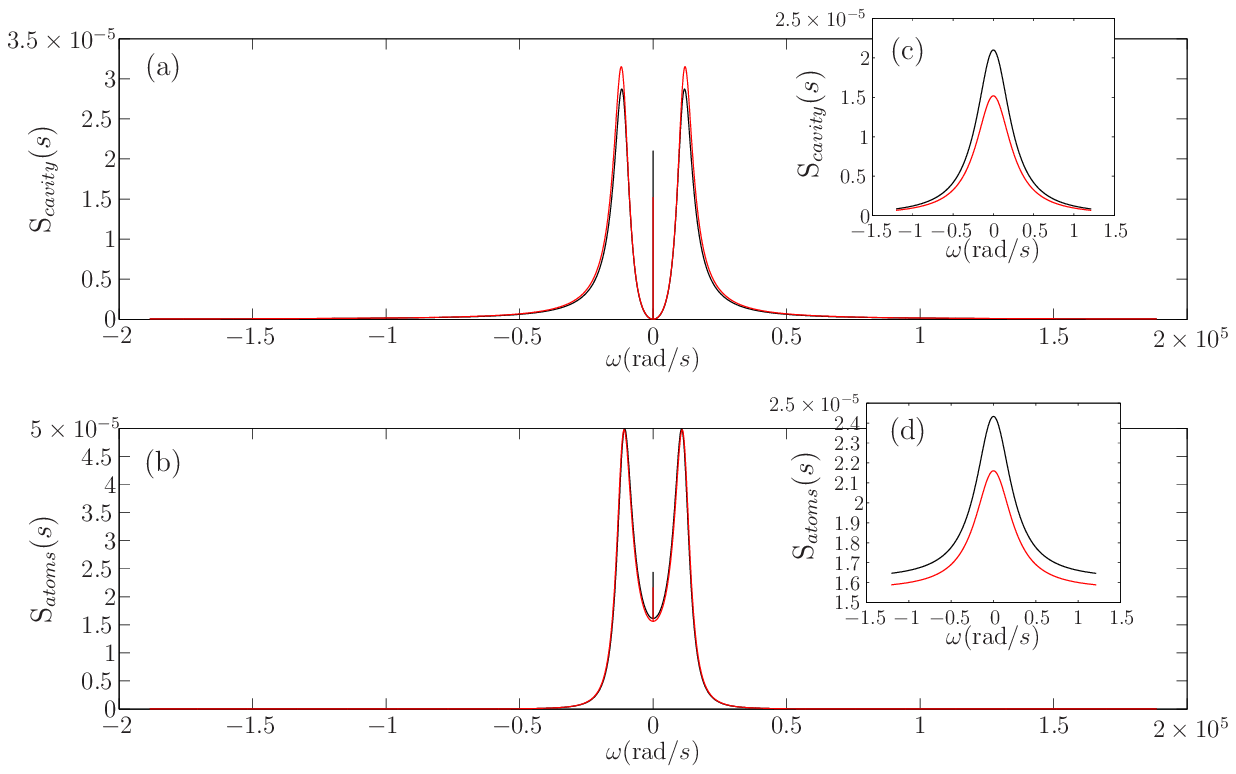}
				\end{center}
				\caption{(Color online) Incoherent spectrum of the mode $\left( \text{a} \right)$
				and of the fluorescence of the ensemble $\left( \text{b} \right)$ for
				$T=\unit{0.025}{\kelvin}$, $\kappa=7\cdot10^{3}$, $\gamma=0.3$, $N=10^{5}$,
				$\eta=9\cdot10^{5}$ in red (gray) and $\eta=10^{6}$ in black. The pump
				driving the system is on resonance with the cavity and the ensemble. Insets
				$\left( \text{c} \right)$ and $\left( \text{d} \right)$ show a magnification of
				the central peak of each spectrum. The lower red (gray) line represents the result
				for $\eta=9\cdot10^{5}$.}
		\label{fig:lateral_end_paper_T0025_N10to5_g40_gamma03_kappa7000_eta_1_10to6_collect}
	\end{figure*}

The spectral intensity distribution of the coherently pumped cavity is calculated in a
way similar to that described in Sec.~\ref{sec:Spectrum}.
In contrast to the incoherent pump process that excites atoms in a noncollective way, as
can be seen from the Liouvillian, the interaction with the coherently pumped mode is a
collective interaction.
To demonstrate this behavior, we calculate the spectral distribution of the mode intensity,
without caring about the reservoir it radiates into, and the spectrum of the fluorescence
of the atoms.\\
The cavity mode and the atomic transition are assumed to be on resonance with
$\omega_{a}=\omega_{m}=2\pi \cdot \unit{6.83}{\giga \hertz}$, which is also the
frequency of the pump laser. To calculate the incoherent part of the spectra we need the
Fourier transform of the two-time correlation functions
\begin{multline}
	\lim_{t\rightarrow \infty}\left< a^{\dagger}\left( t
	\right)a\left( t+\tau \right) \right>_{c}=\\
	\lim_{t\rightarrow \infty}\left( \left< a^{\dagger}\left( t
	\right)a\left( t+\tau \right) \right> -\left< a^{\dagger }(t) \right>\left< a\left(
	t+\tau \right) \right> \right)
		\label{eq:inc_two_time1}
\end{multline}
and
\begin{multline}
	\lim_{t\rightarrow \infty}\left< \sigma^{+}_{i}\left( t
	\right)\sigma^{-}_{j}\left( t+\tau \right) \right>_{c}=\\
	\lim_{t\rightarrow \infty}\left(
	\left< \sigma^{+}_{i}\left( t
	\right)\sigma^{-}_{j}\left( t+\tau \right) \right> -\left< \sigma^{+}_{i}(t) \right>\left<
	\sigma^{-}_{j}\left(
	t+\tau \right) \right> \right)\ .
	\label{eq:inc_two_time2}
\end{multline}
The quantum regression theorem and Eqs.~\eqref{eq:einer1} to \eqref{eq:einer5} give
\begin{multline}
	\frac{\mathrm{d}}{\mathrm{d}\tau}\left< a^{\dagger }\left( 0 \right)a\left( \tau
	\right) \right>_{c}=\\
	-\left( \kappa+\mathrm{i}\Delta_{m} \right)\left< a^{\dagger }\left( 0 \right)a\left( \tau
	\right) \right>_{c}-\mathrm{i}gN\left< a^{\dagger }\left( 0 \right)\sigma^{-}_{i}\left( \tau
	\right) \right>_{c}
	\label{eq:diff_two_time1}
\end{multline}
and
\begin{multline}
	\frac{\mathrm{d}}{\mathrm{d}\tau}\left< \sigma^{+}_{i}\left( 0
	\right)\sigma^{-}_{j}\left( \tau
	\right)\right>_{c}=\\
	-\left( \frac{\gamma_{a}}{2}+\mathrm{i}\Delta_{a} \right)\left<
	\sigma^{+}_{i}\left( 0 \right)\sigma^{-}_{j}\left( \tau
	\right)\right>_{c}\\
	+\mathrm{i}g\left( \left< a\left( \tau \right) \right>\left<
	\sigma^{+}_{i}\left( 0 \right)\sigma^{z}_{j}\left( \tau \right)\right>_{c}+\left<
	\sigma^{z}_{i}\left( \tau
	\right) \right>\left< \sigma^{+}_{i}\left( 0 \right)a\left( \tau \right) \right>_{c}
	 \right)
	\label{eq:diff_two_time2}
\end{multline}

where we use $\lim_{t\rightarrow \infty}\left< a^{\dagger}\left( t
	\right)a\left( t+\tau \right) \right>_{c}\equiv \left< a^{\dagger}\left( 0
	\right)a\left( \tau \right) \right>_{c}$ and $\lim_{t\rightarrow \infty}\left<
	\sigma^{+}_{i}\left( t
	\right)\sigma^{-}_{j}\left( t+\tau \right) \right>_{c}\equiv \left<
	\sigma^{+}_{i}\left( 0
	\right)\sigma^{-}_{j}\left( \tau \right) \right>_{c}$. Equations~\eqref{eq:diff_two_time1}
	and \eqref{eq:diff_two_time2} couple to four other two-time correlation functions that
	have to be calculated. To solve for the desired quantities, we Laplace transform both
	sets of equations and use Cramer's rule to obtain $\widetilde{\left< a^{\dagger}\left( 0
	\right)a\left( \tau \right) \right>_{c}}\left( s \right)$ and $\widetilde{\left<
	\sigma_{i}^{+}\left( 0
	\right)\sigma^{-}_{j}\left( \tau \right) \right>_{c}}\left( s \right)$. The necessary
	steady-state
	values are obtained numerically.\par

The incoherent spectra of the mode and of the atoms both show a narrow peak at
$\omega_{a}=\omega_{m}$ which has a width of $\approx 2 \gamma_{a}$ (see
Fig.~\ref{fig:lateral_end_paper_T0025_N10to5_g40_gamma03_kappa7000_eta_1_10to6_collect}).
The double-peaked structure is a remainder of thermal excitations acting as a broad band probe for
the ensemble-cavity system. With increasing strength of the coherent pump the narrow
central peak becomes dominant.  The appearance of the central peak is probably related to
weak contributions from almost-dark states (very weakly coupled to the mode). Let us
mention in this context that the atomic ensemble is not restricted to a manifold of the
Dicke states with fixed $J$ since we include spontaneous emission in our model. It is
hence possible that the ensemble ends up in a dark state, where it does not couple to the
cavity mode. The time-constant that determines the decay into and the decay of such a dark
state is of the order $1/\gamma_{a}$. This allows for the buildup of long time coherences,
and the times the ensemble is in a dark state significantly change the statistics of the
photon emission. The result is then a narrow peak in the incoherent
spectrum~\citep{plenio1998quantum,zoller1987quantum}, where the width of the peak is
determined by the characteristic time the ensemble resides in a bright or dark state, in
our case $\gamma_{a}$.\\
In the case of an incoherently pumped ensemble, the narrow peak does not arise. A reason
for this can be the nature of the incoherent pump which is noncollective and hence able
to pump the ensemble out of a dark state in a shorter time. This is not possible in the
case of the coherently pumped cavity: Spontaneous emission brings the ensemble to a dark
state, but the collective interaction with the mode cannot reach it there.

%\clearpage

%%%%%%%%%%%%%%%%%%%%%%%%%%%%%%%%%%%%%%%%%%%%%%%%%%%%%%%%%%%%%%%%%%%%%%%%%%%%%%%%%%%%%%%%%%%%%%%%%%%%%%%%%%%%%%%%%%

\section{Superradiance}\label{sec:superradiance}

\begin{figure}[!]
	\begin{center}
		\includegraphics[width=0.5\textwidth]{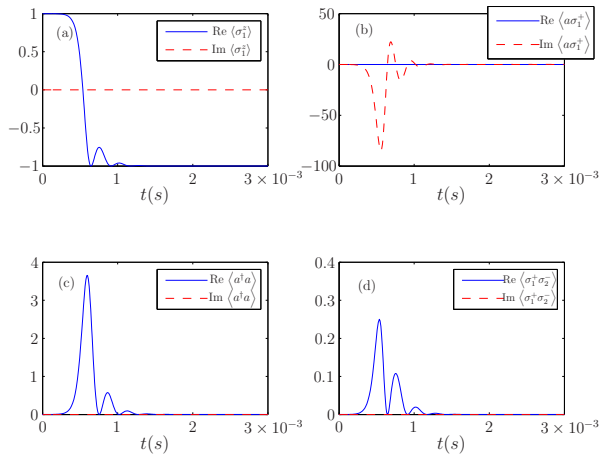}
			\end{center}
	\caption{(Color online) Dynamics of superradiant emission: numerical solutions of the dynamical
	equations for an ensemble of $N=10^{5}$ atoms. The rapid drop of the inversion $\left<
	\sigma_{1}^{z} \right>$ during the emission can be seen in $\left( \text{a} \right)$.
	The exchange of excitations between the ensemble and the cavity is characterized by
	$\left< a\sigma_{1}^{+} \right>$ in $\left( \text{b} \right)$. Negative imaginary part
	of $\left< a\sigma_{1}^{+} \right>$ indicates emission from the ensemble into the
	cavity, where a positive imaginary part indicates absorption of cavity photons by the
	ensemble. In $\left( \text{c} \right)$ the number of photons in the cavity is depicted.
	Panel $\left( \text{d} \right)$ shows the spin-spin correlation $\left<
	\sigma_{1}^{+}\sigma_{2}^{-} \right>$. In this example the temperature of the mode was
	chosen to be $T=\unit{4}{\kelvin}$.}
	\label{fig:superradiant_decay_dynamics_paper}
\end{figure}

\begin{figure}[h!]
	\begin{center}
		\includegraphics[width=0.4\textwidth]{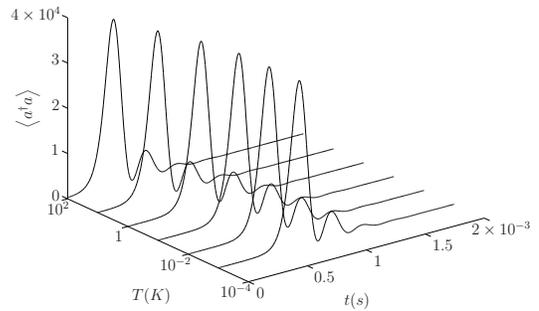}
	\end{center}
	\caption{Dynamics of the photon number in the cavity with increasing temperature. The
	onset of the superradiant emission is shifted to earlier times since the initially
	present thermal photons contribute to the fluctuations that trigger the emission
	process.}
	\label{fig:superradiance_vs_T_kappa_7_10to3_g40_gamma03_N10to5}
\end{figure}

A great advantage of the considered setup is that one has full control of the atomic
state. Hence, instead of starting at a zero-temperature ground state we can prepare an
almost fully inverted ensemble, which can feed energy into the system and corresponds to
an effective negative temperature~\citep{gardiner1991quantum}. Since we have no initial
phase bias in the system, Eqs.~\eqref{eq:sz}--\eqref{eq:spsm1} are suitable for describing
the dynamics. The resulting superradiant dynamics for an initially fully inverted ensemble
is depicted in Fig.~\ref{fig:superradiant_decay_dynamics_paper}. In free space the
emission occurs in a characteristic burst of duration $\approx
\frac{1}{\gamma_{a}N}$~\citep{gross1982set}. The presence of the cavity causes a partial
reabsorption of the emitted photons, which can be seen in
Fig.~\ref{fig:superradiant_decay_dynamics_paper} (c). Due to the large number of emitted
photons it should be clearly detectable even on a fairly high thermal background.
Following the pulse shape, one also can extract the effective coupling parameters to
characterize the system.

The process of superradiance can create a transient entangled state of the
ensemble~\citep{milman2006witnessing}. This entanglement can be revealed by entanglement
witnesses which can be inferred from the calculated observables. We have seen some
indication of such entanglement appearing. However, the persistence of the entanglement
under the influence of noise and with the presence of the cavity will be part of future
work.

The onset of superradiant emission is determined by spontaneously emitted photons that
trigger the forthcoming burst of radiation. The presence of thermal photons is expected to
reduce the time until the onset of the burst. This behavior is recovered by our equations
as shown in Fig.~\ref{fig:superradiance_vs_T_kappa_7_10to3_g40_gamma03_N10to5}, where we
depict the dynamics of the photon number in the cavity for different temperatures.

%  more comments on thermal initiation of superradint pulse ...

%\clearpage

%%%%%%%%%%%%%%%%%%%%%%%%%%%%%%%%%%%%%%%%%%%%%%%%%%%%%%%%%%%%%%%%%%%%%%%%%%%%%%%%%%%%%%%%%%%%%%%%%%%%%%%%%%%%%%%%%%%%%%%%

\section{Narrowbandwidth hyperfine micromaser }\label{sec:laser}

\begin{figure*}[!]
	\begin{center}
		\includegraphics{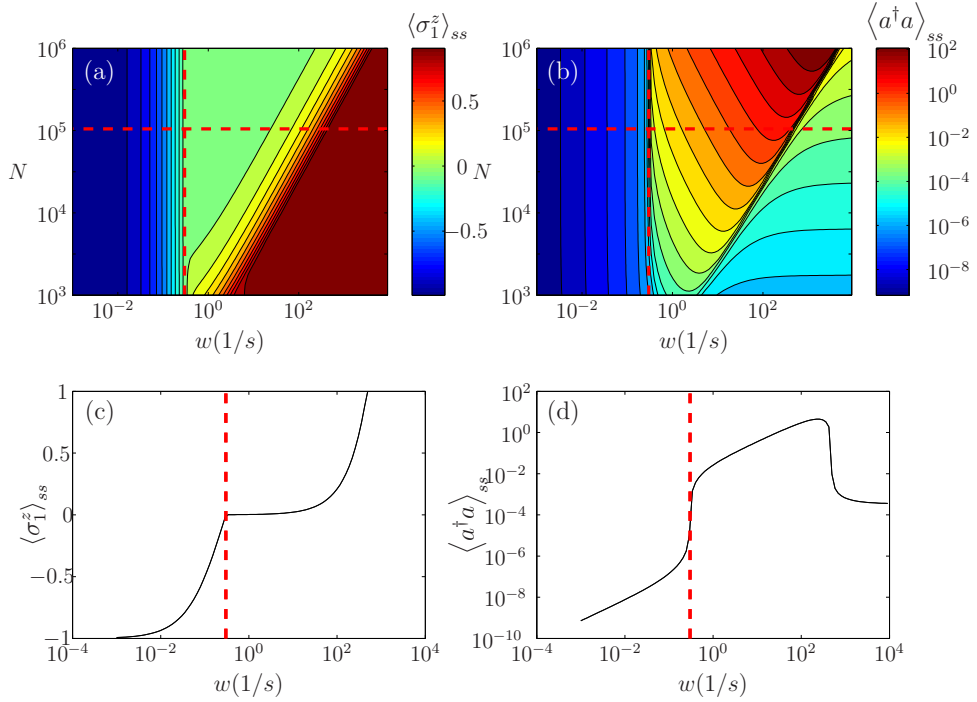}
			\end{center}
	\caption{(Color online) Steady-state inversion of the ensemble $\left( \text{a}  \right)$ and
	occupation of the cavity mode $\left( \text{b}  \right)$ for varying ensemble size $N$
	and pump strength $w$. The temperature was chosen to be $T_{1}=\unit{0.001}{\kelvin}$, which
	corresponds to an empty cavity ($\bar{n}=0$). Vertical red dashed lines mark the masing threshold
	$w=\gamma_{a}=0.3$. The horizontal red dashed lines at $N=10^{5}$ indicate the position of the curves
	shown in $\left( \text{c}  \right)$ and $\left( \text{d}  \right)$.  From
	$\left( \text{c}  \right)$ we recover the passage of the inversion through zero for
	$w=\gamma_{a}=0.3$. At this point we see in $\left( \text{d}  \right)$ a rapid increase of
	the photon number in the mode.}
	\label{fig:linewidth_paper_1_power_and_inversion}
\end{figure*}

The collectively coupled ensemble can be used to construct a stripline micromaser with a
very low linewidth. To this aim the inversion of the ensemble is sustained by an external
incoherent pump of the atoms. In contrast to the calculations in Sec.~\ref{sec:Spectrum}, we
ignore the fact that the cavity radiates into a thermally occupied reservoir. After
passing the masing threshold, the thermal occupation outside becomes negligible.  To
determine the linewidth of the emitted light we calculate the Laplace transform of the
two-time correlation function $\left< a^{\dagger }(t)a(0) \right>$. Using the quantum
regression theorem we find
\begin{multline}
	\frac{\mathrm{d}}{\mathrm{d}t}\begin{pmatrix}
		\left< a^{\dagger }(t)a(0) \right>\\
	 \left< \sigma_1^{+}(t)a(0) \right>
	\end{pmatrix}=\\
	\begin{pmatrix}
		-\kappa & \mathrm{i}gN\\
		-\mathrm{i}g\left< \sigma_1^{z} \right>_{ss}& -\frac{w+\gamma_{a}}{2}
	\end{pmatrix}\cdot
	\begin{pmatrix}
			\left< a^{\dagger }(t)a(0) \right>\\
	 \left< \sigma_1^{+}(t)a(0) \right>
	\end{pmatrix}\ .
	\label{eq:sys}
\end{multline}
Laplace Transform of Eq.~\eqref{eq:sys} gives
\begin{align}
        \begin{pmatrix}
                \kappa+s & -\mathrm{i}gN\\
                \mathrm{i}g\left< \sigma_1^{z} \right>_{ss} & \frac{w+\gamma_{a}}{2}+s
        \end{pmatrix}\cdot
        \begin{pmatrix}
                \widetilde{\left< a^{\dagger }(t)a(0) \right>}\\
                \widetilde{\left< \sigma_1^{+}(t)a(0) \right>}
        \end{pmatrix}=
        \begin{pmatrix}
                \left< a^{\dagger }a \right>_{ss}\\
                \left< \sigma_1^{+}a \right>_{ss}
        \end{pmatrix}\ ,
        \label{eq:laplace}
\end{align}
where $\left< \cdot \right>_{ss}$ denotes steady state values and $\widetilde{ \cdot }$
denotes Laplace transformed quantities.\par
\begin{figure*}[!]
	\begin{center}
		\includegraphics{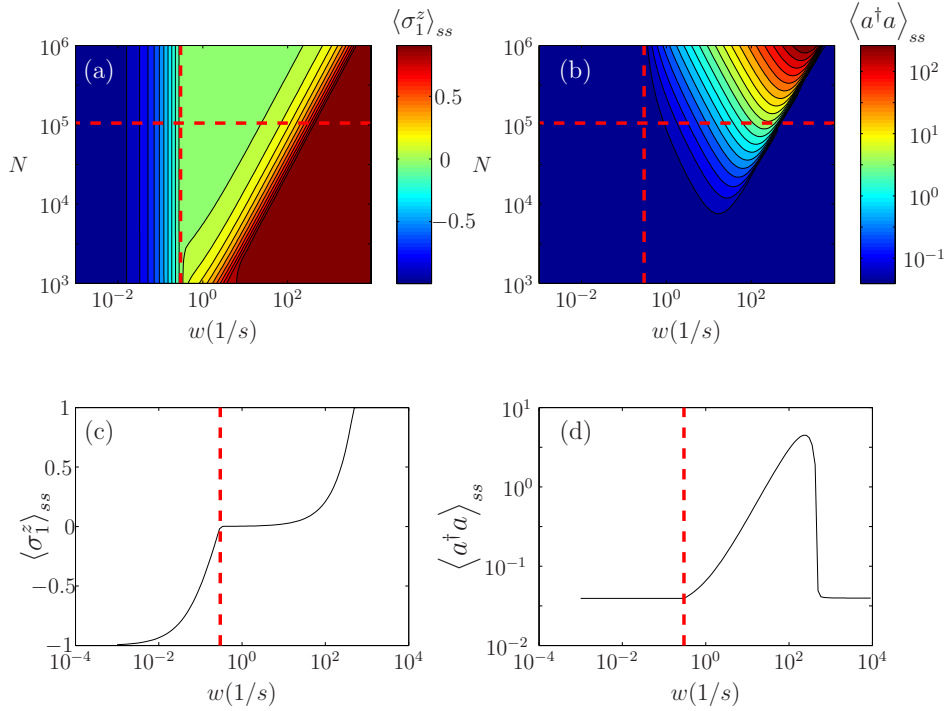}
			\end{center}
	\caption{(Color online) Same as Fig.~\ref{fig:linewidth_paper_1_power_and_inversion} with
	$T_{2}=0.1$, which corresponds to $\bar{n}=0.04$. In $\left( \text{d}  \right)$ the
	thermal excitations in the mode appear as a constant background from which the increase
	due to the pump stands out.}
	\label{fig:linewidth_paper_2_power_and_inversion}
\end{figure*}

 The steady-state values on the right-hand side
of Eq.~\eqref{eq:laplace} can be obtained analytically. Setting the time derivatives of
the dynamical equations to zero, a quadratic equation for $\left< \sigma_{1}^{z}
\right>_{ss}$ is attained. One of the solutions yields a physically meaningful result for
calculating the remaining steady-state values and hence $\left< a^{\dagger }a \right>_{ss}$
and $\left< \sigma_1^{+}a \right>_{ss}$.  To illustrate the effect of the increasing pump
strength, we show the steady-state inversion of the ensemble and the occupation of the
cavity in Figs.~\ref{fig:linewidth_paper_1_power_and_inversion} and ~\ref{fig:linewidth_paper_2_power_and_inversion}. Once the critical pump strength is
reached, the systems behave identically for different temperatures. The number of atoms in the
ensemble is varied between $10^{3}$ and $10^{6}$, whereas the pump parameter $w$ ranges
from $10^{-3}$ to $10^{4}$. In both figures we mark the pump strength $w=\gamma_{a}=0.3$,
for which we find the inversion becomes positive, with a vertical red dashed line. At
this point we also find a rapid increase of the number of photons in the cavity. The
horizontal lines mark the cross sections for $N=10^{5}$ shown in what follows.\par

\begin{figure}[!]
\centering
\subfigure{
\includegraphics[width=0.5\textwidth]{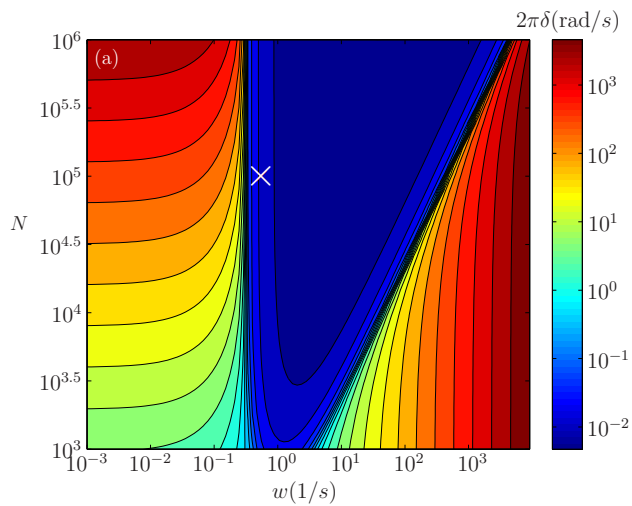}
}
\subfigure{
\includegraphics[width=0.5\textwidth]{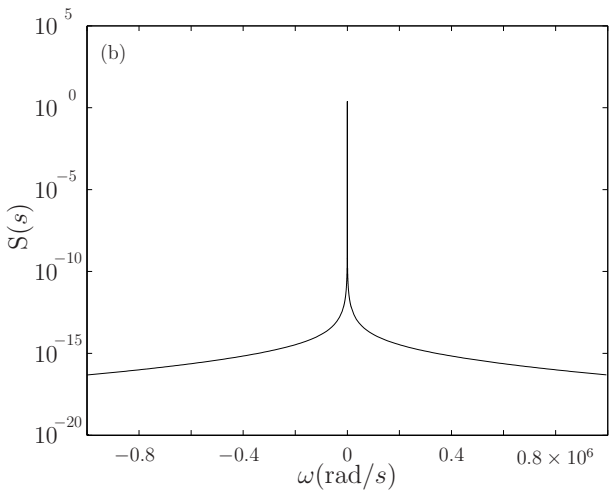}
}
\caption{(Color online) $\left( \text{a} \right)$ Linewidth of the spectrum
$\operatorname{S}\left( \omega \right)$. For each set of parameters we numerically
determine the linewidth of the spectrum. The parameters were chosen to be
$\kappa=7\cdot10^5$, $\gamma_{a}=0.3$, $g=40$,
$\omega_{m}=\omega_{a}=2\pi\cdot6.83\cdot10^9$, $T=0.001$. $\left( \text{b} \right)$
Exemplary spectrum for $N=10^5$ and $w=0.55$, marked in $\left( \text{a} \right)$ by the
white cross.}
\label{fig:linewidth_paper_1_flat}
\end{figure}

\begin{figure}[!]
	\centering
\subfigure{
\includegraphics[width=0.5\textwidth]{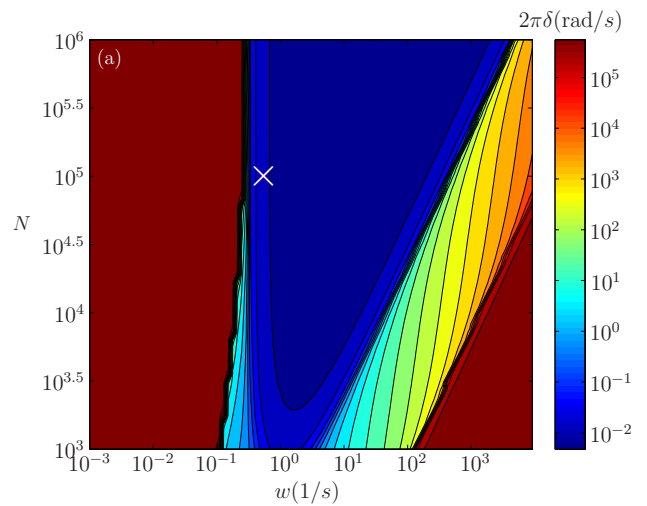}
}
\subfigure{
\includegraphics[width=0.5\textwidth]{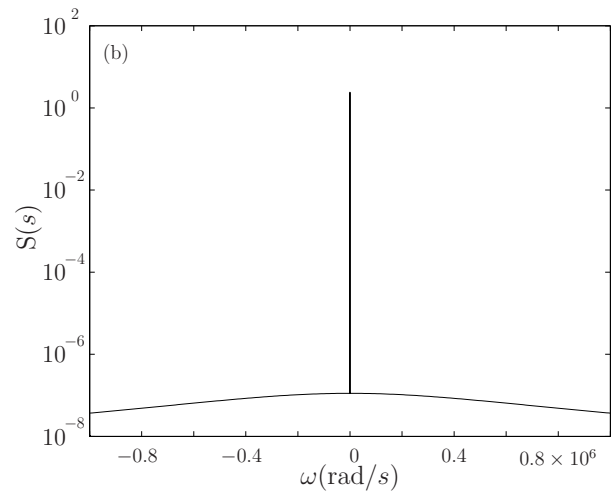}
}
\caption{(Color online) Finite temperature effects in the spectrum. Panel $\left( \text{a}
\right)$ again shows the linewidth of the spectrum. Below the critical pump strength we
recover the linewidth of the cavity. The parameters were chosen to be $\kappa=7\cdot10^5$,
$\gamma_{a}=0.3$, $g=40$, $\omega_g=\omega_{a}=2\pi\cdot6.83\cdot10^9$, $T=0.1$. $\left(
\text{b} \right)$: Exemplary spectrum for $N=10^5$ and $w=0.55$, marked in $\left(
\text{a} \right)$ by the white cross.}
\label{fig:linewidth_paper_2_flat}
\end{figure}

To solve for $\widetilde{\left< a^{\dagger }(t)a(0) \right>}$, we use Cramers rule on
Eq.~\eqref{eq:laplace}, which yields
\begin{align}
        \widetilde{\left< a^{\dagger }(t)a(0) \right>}(s)=
        \frac{\left< a^{\dagger }a \right>_{ss}\left( \frac{w+\gamma_{a}}{2}+s
        \right)-\mathrm{i}gN\left< \sigma_1^{+}a
        \right>_{ss}}{(\kappa+s)(\frac{w+\gamma_{a}}{2}+s)-g^2N\left< \sigma_1^{z}
				\right>_{ss}}\ ,
        \label{eq:det_spec}
\end{align}
so that with $s=-\mathrm{i}\omega$ the spectrum is given by
\begin{align}
	\operatorname{S}\left( \omega
	\right)&=\frac{1}{2\pi}\left(\widetilde{\left< a^{\dagger }(t)a(0) \right>}\left( \omega \right)+\widetilde{\left< a^{\dagger
}(t)a(0) \right>}^{*}\left( \omega \right)\right)\ .
	\label{eq:spec}
\end{align}
For each set of parameters we calculate the spectrum and determine the linewidth
numerically. The linewidth of the maser for two different temperatures
$T_{1}=\unit{0.001}{\kelvin}$ and $T_{2}=\unit{0.1}{\kelvin}$ is shown in
Figs.~\ref{fig:linewidth_paper_1_flat}$\left( \text{a} \right)$ and ~\ref{fig:linewidth_paper_2_flat}$\left( \text{a} \right)$. For $w=\gamma_{a}=0.3$ we
see a rapid drop in the linewidth for both temperatures and a resulting minimal linewidth
of $\delta=\unit{\frac{1}{2\pi}4.7\cdot10^{-3}}{\hertz}$. Above the critical pump strength
the pump noise destroys the coherence between the individual atoms~\citep{meiser2009pml}.
In in Figs.~\ref{fig:linewidth_paper_1_flat}$\left( \text{b} \right)$ and ~\ref{fig:linewidth_paper_2_flat}$\left( \text{b} \right)$ we plot exemplary spectra
for $N=10^{5}$ and $w=0.55$, marked by the white cross.\par

%\clearpage

For $T_{2}=\unit{0.1}{\kelvin}$ the cavity contains on average $\bar{n}=0.04$ photons
that can be recovered from the constant background in
Figs.~\ref{fig:linewidth_paper_2_power_and_inversion}$\left( \text{b}  \right)$ and ~\ref{fig:linewidth_paper_2_power_and_inversion}$\left(
\text{d}  \right)$. Since the number of thermal photons is small compared to the
considered ensembles, the inversion is nonsensitive to the increased temperature.

In Fig.~\ref{fig:linewidth_paper_2_flat} we recover the linewidth of the cavity
$\kappa=7\cdot10^{5}$ if the pump is below threshold and again if the pump exceeds a
critical strength $w_{\text{max}}$. Above $w_{\text{max}}$ the coherence between different
spins is destroyed by the pump noise~\citep{meiser2009pml}. The behavior between the
threshold and $w_{\text{max}}$ resembles the behavior for $T=\unit{0.001}{\kelvin}$ shown
in Fig.~\ref{fig:linewidth_paper_1_flat}.

%\clearpage
\section{Conclusions}
Our studies show that a hybrid cavity QED system consisting of a stripline microwave
resonator at finite $T$ and an ensemble of ultracold atoms is a rich and versatile setup
for
observing and testing prominent quantum physics phenomena. The effectively very cold
temperature and good localization of the atomic cloud allow symmetric collective strong
coupling to the microwave mode. While the weak magnetic dipole coupling requires large
atom numbers and an extremely well localized microwave mode to obtain significant
coupling, it also makes the system quite immune to external noise. In addition to the
long lifetime of the atomic states, this renders the system an ideal quantum memory or
allows for very narrow spectral response or gain. As all the atoms are identical and well
trapped, the system exhibits only a very narrow inhomogeneous broadening. Operated in an
active way, one thus can envisage a truly microscopic maser with an very narrow linewidth
directly locked to an atomic clock transition. The uniform coupling and the possibility of
efficient optical pumping enables the study of superradiant decay into the stripline mode,
where a precise phase and intensity analysis of the emitted radiation can be performed.

While many of our considerations are guided by parameters expected from an ultracold atom
ensemble, it is easy to generalize to alternative setups using NV-centers or other solid-state ensembles. There larger ensembles can be easily envisaged but one also gets
much more varying coupling constants and larger inhomogeneous widths. It is not obvious
whether the technically more simple setup and larger numbers in this case can compensate
for these imperfections.

This could be particularly important for a next step: possible optical readout of the
ensemble. For the atomic case, the uniformity of the coupling over many optical wavelengths
should allow a nice directional readout of the ensemble state, once a laser could coupled
in.

\begin{acknowledgments}
This work was supported by a DOC-fFORTE-fellowship of the Austrian Academy of Sciences and
the European Union project MIDAS.
\end{acknowledgments}

\clearpage

%%%%%%%%%%%%%%%%%%%%%%%%%%%%%%%%%%%%%%%%%%%%%%%%%%%%%%%%%%%%%%%%%%%%%%%%%%%%%%%%%%%%%%%%%%%%%%%%%%%%%%%%%%%%%%%%%%%%%%

\appendix
\section{Coupled Equations} \label{app:A}
Transformation to a rotating frame with respect to the frequency of the pump $\omega_{l}$
results in $\Delta_{m}=\omega_{m}-\omega_{l}$ for the detuning of the cavity and
$\Delta_{a}=\omega_{a}-\omega_{l}$ for the detuning of the atoms. The coupled equations
are given by: 

\begin{widetext}
\begin{align}
	\frac{\mathrm{d}}{\mathrm{d}t}\left< a \right>&=-\left( \kappa+\mathrm{i}\Delta_{m} \right)\left< a
	\right>-\mathrm{i}gN\left< \sigma_{1}^{-} \right>+\eta\label{eq:einer1}\\
	\frac{\mathrm{d}}{\mathrm{d}t}\left< a^{\dagger } \right>&=-\left( \kappa-\mathrm{i}\Delta_{m}
	\right)\left< a^{\dagger }
	\right>+\mathrm{i}gN\left< \sigma_{1}^{+} \right>+\eta^{*}\label{eq:einer2}\\
	\frac{\mathrm{d}}{\mathrm{d}t}\left< \sigma_{1}^{z} \right>&=-2\mathrm{i}g\left( \left( \left<
	\sigma_{1}^{+}a \right>_c+\left< \sigma_{1}^{+} \right>\left< a \right> \right)-\left(
	\left< \sigma_{1}^{-}a^{\dagger } \right>_c+\left< \sigma_{1}^{-} \right>\left<
	a^{\dagger }
	\right> \right) \right)\nonumber\\
	&-\gamma_{a}\left( 1+\left< \sigma_{1}^{z} \right> \right)\label{eq:einer3}\\
	\frac{\mathrm{d}}{\mathrm{d}t}\left< \sigma_{1}^{-} \right>&=-\left( \frac{\gamma_{a}}{2}+\mathrm{i}\Delta_a
	\right)\left< \sigma_{1}^{-} \right>+\mathrm{i}g\left( \left< \sigma_{1}^{z}a
	\right>_c+\left< \sigma_{1}^{z} \right>\left< a \right> \right)\label{eq:einer4}\\
	\frac{\mathrm{d}}{\mathrm{d}t}\left< \sigma_{1}^{+} \right>&=-\left( \frac{\gamma_{a}}{2}-\mathrm{i}\Delta_a
	\right)\left< \sigma_{1}^{+} \right>-\mathrm{i}g\left( \left< \sigma_{1}^{z}a^{\dagger }
	\right>_c+\left< \sigma_{1}^{z} \right>\left< a^{\dagger } \right> \right)
	\label{eq:einer5}
\end{align}

\begin{align}
	\frac{\mathrm{d}}{\mathrm{d}t}\left(\left< a \sigma_{1}^{+}
	\right>_c+\left< a \right>\left< \sigma_{1}^{+} \right>\right)&=-\left( \kappa+\frac{\gamma_{a}}{2}+\mathrm{i}\left(
	\Delta_{m}-\Delta_a \right) \right)\left(\left< a \sigma_{1}^{+}
	\right>_c+\left< a \right>\left< \sigma_{1}^{+} \right>\right)\nonumber\\
	&-\mathrm{i}g\left( \frac{\left< \sigma_{1}^{z} \right>+1}{2}+\left( N-1
	\right)\left(\left< \sigma_{1}^{+}\sigma_{2}^{-} \right>_c+\left< \sigma_{1}^{+}
	\right>\left< \sigma_{2}^{-} \right>\right) \right)\nonumber\\
	&-\mathrm{i}g \left< \sigma_{1}^{z}a^{\dagger}a \right>+\eta\left< \sigma_{1}^{+} \right>
	\label{eq:asplus}
\end{align}

\begin{align}
	\frac{\mathrm{d}}{\mathrm{d}t}\left(\left< a \sigma_{1}^{z} \right>_c+\left< \sigma_{1}^{z} \right>\left< a
	\right>\right)&=-\left( \kappa+\mathrm{i}\Delta_{m}
	\right)\left(\left< a \sigma_{1}^{z} \right>_c+\left< \sigma_{1}^{z} \right>\left< a
	\right>\right)+\eta\left< \sigma_{1}^{z} \right>\nonumber\\
	&-2\gamma_{a}\left(\left< a\sigma_{1}^{+} \right>_c\left< \sigma_{1}^{-}
	\right>+\left< a\sigma_{1}^{-} \right>_c\left< \sigma_{1}^{+} \right>+\left< a \right>\left<
	\sigma_{1}^{+}\sigma_{1}^{-} \right>\right)\notag\\
	&-
	\begin{multlined}[t][0.8\linewidth]
		\mathrm{i}g\left[2\left( 2\left< \sigma_{1}^{+}a \right>_c\left< a \right>+\left< aa
		\right>_c\left< \sigma_{1}^{+} \right>+\left< a
		\right>\left< a \right>\left< \sigma_{1}^{+} \right> \right)\right.\\
		\shoveleft[1cm]-\left. \left( 1+2\left<
		a^{\dagger }a \right>_c
		\right)\left< \sigma_{1}^{-} \right>-2\left< a\sigma_{1}^{-} \right>_c\left<
		a^{\dagger }
		\right>\right.\\
		\shoveleft[1.2cm]\left.-2\left< a^{\dagger }\sigma_{1}^{-} \right>_c\left< a \right>-2\left<
		a^{\dagger }
		\right>\left< a \right>\left< \sigma_{1}^{-} \right>\right.\\
		\shoveleft[1.2cm]\left.+\left( N-1 \right)\left( \left< \sigma_{1}^{z}
		\sigma_{2}^{-}
		\right>_c+\left< \sigma_{1}^{z} \right>\left< \sigma_{2}^{-} \right> \right)\right]
	\end{multlined}
	\label{eq:asz}
\end{align}

\begin{align}
	\frac{\mathrm{d}}{\mathrm{d}t}\left(\left<\sigma_{1}^{+}\sigma_{2}^{-}\right>_c+\left< \sigma_{1}^{+}
	\right>\left< \sigma_{2}^{+} \right>\right)&=-\mathrm{i}g\left[ \left<
	\sigma_{1}^{z}\sigma_{2}^{-} \right>_c\left< a^{\dagger } \right>+\left<
	\sigma_{1}^{z}a^{\dagger } \right>_c\left< \sigma_{1}^{-} \right>+\left<
	\sigma_{1}^{-}a^{\dagger }
	\right>_c\left< \sigma_{1}^{z} \right>+\left< \sigma_{1}^{-} \right>\left<
	\sigma_{1}^{z}
	\right>\left< a^{\dagger } \right>\right.\notag\\
	&-\left.\left( \left<
	\sigma_{1}^{z}\sigma_{2}^{+} \right>_c\left< a \right>+\left<
	\sigma_{1}^{z}a  \right>_c\left< \sigma_{1}^{+} \right>+\left<
	\sigma_{1}^{+}a
	\right>_c\left< \sigma_{1}^{z} \right>+\left< \sigma_{1}^{+} \right>\left<
	\sigma_{1}^{z}
	\right>\left< a \right> \right)
	\right]\notag\\
	&-\gamma_{a}\left(\left<\sigma_{1}^{+}\sigma_{2}^{-}\right>_c+\left< \sigma_{1}^{+}
	\right>\left< \sigma_{2}^{-} \right>\right)
	\label{eq:spsm}
\end{align}

\begin{align}
	\frac{\mathrm{d}}{\mathrm{d}t}\left(\left< a^{\dagger }a \right>_c+\left< a^{\dagger } \right>\left< a
	\right>\right)=&-\mathrm{i}gN\left( \left(\left< a^{\dagger }
	\sigma_{1}^{-}
	\right>_c+\left< a^{\dagger } \right>\left< \sigma_{1}^{-} \right>\right)-\left(\left< a \sigma_{1}^{+}
	\right>_c+\left< a \right>\left< \sigma_{1}^{+} \right>\right) \right)\notag\\
	&-2\kappa\left(\left< a^{\dagger }a \right>_c+\left< a^{\dagger } \right>\left< a
	\right>\right)+2\kappa\bar{n}+ \eta^{*}\left< a \right>+\eta\left< a^{\dagger }
	\right>
	\label{eq:adaggera}
\end{align}

\begin{align}
	\frac{\mathrm{d}}{\mathrm{d}t}\left(\left< a \sigma_{1}^{-} \right>_c+\left< a \right>\left<
	\sigma_{1}^{-} \right>\right)&=-\left( \mathrm{i}\left(
	\Delta_{m}+\Delta_a \right)+\kappa+\frac{\gamma_{a}}{2} \right)\left(\left< a \sigma_{1}^{-} \right>_c+\left< a \right>\left<
	\sigma_{1}^{-} \right>\right)\notag\\
	&-\mathrm{i}g(N-1)\left(\left< \sigma_{1}^{-}\sigma_{2}^{-}
	\right>_c+\left< \sigma_{1}^{-} \right>\left<\sigma_{1}^{-}
	\right>\right)+\eta\left< \sigma_{1}^{-} \right>\nonumber\\
	&+\mathrm{i}g\left( 2\left< \sigma_{1}^{z}a \right>_c\left<a  \right>+\left< aa
	\right>_c\left< \sigma_{1}^{z} \right>+\left< a \right>\left< a \right>\left<
	\sigma_{1}^{z}
	\right>\right)
	\label{eq:as-2}
\end{align}

\begin{align}
	\frac{\mathrm{d}}{\mathrm{d}t}\left( \left< a^{\dagger }a^{\dagger } \right>_c-\left< a^{\dagger } \right>\left<
	a^{\dagger } \right> \right)&=-\left(2\kappa-\mathrm{i}2\Delta_{m}\right)\left( \left< a^{\dagger
	}a^{\dagger } \right>_c+\left< a^{\dagger } \right>\left< a^{\dagger } \right>
	\right)\nonumber
	\\&+2\mathrm{i}gN\left(\left< \sigma_{1}^{+}a^{\dagger } \right>_c+\left< a^{\dagger }
	\right>\left< \sigma_{1}^{+} \right>\right)+2\eta^{*}\left< a^{\dagger } \right>
	\label{eq:adaggeradagger}
\end{align}

\begin{align}
	\frac{\mathrm{d}}{\mathrm{d}t}\left( \left< \sigma_{1}^{-}\sigma_{2}^{-}\right>_{c} +\left< \sigma_{1}^{-}
	\right>\left< \sigma_{1}^{-} \right> \right)&=-2\left(
	\frac{\gamma_{a}}{2}+\mathrm{i}\Delta_a \right)\left( \left< \sigma_{1}^{-}\sigma_{2}^{-}
	\right>_c+\left< \sigma_{1}^{-}
	\right>\left< \sigma_{2}^{-} \right> \right)\notag\\
	&+\mathrm{i}2g\left( \left< \sigma_{1}^{z}\sigma_{2}^{-} \right>_c\left< a \right>+\left<
	\sigma_{1}^{z}a \right>_c\left< \sigma_{1}^{-} \right>+\left< \sigma_{1}^{-}a
	\right>_c\left< \sigma_{1}^{z} \right>+\left< \sigma_{1}^{z} \right>\left<
	\sigma_{1}^{-}\right>\left< a \right>\right)
	\label{eq:smsm}
\end{align}

\begin{align}
	\frac{\mathrm{d}}{\mathrm{d}t}\left(\left<  \sigma_{1}^{z} \sigma_{2}^{+} \right>_c+\left< \sigma_{1}^{z}
	\right>\left< \sigma_{1}^{+}
	\right>\right)&=\left( -\frac{\gamma_{a}}{2}+\mathrm{i}\Delta_{a}
	\right)\left(\left<  \sigma_{1}^{z} \sigma_{2}^{+} \right>_c+\left< \sigma_{1}^{z}
	\right>\left< \sigma_{2}^{+}
	\right>\right)\nonumber\\
	&-2\gamma_{a}\left(\left< \sigma_{1}^{+}\sigma_{1}^{-} \right>\left< \sigma_{1}^{+}
	\right>+\left< \sigma_{1}^{+}\sigma_{2}^{+} \right>_c\left< \sigma_{1}^{-}
	\right>+\left< \sigma_{1}^{-}\sigma_{2}^{+} \right>_c\left< \sigma_{1}^{+}
	\right>\right)\notag\\
	&-
	\begin{multlined}[t][0.8\linewidth]
		\mathrm{i}g\left[2\left( \left< \sigma_{1}^{+}\sigma_{2}^{+} \right>_c\left< a
		\right>+2\left< \sigma_{1}^{+}a \right>_c\left< \sigma_{1}^{+} \right>+\left<
		\sigma_{1}^{+}
		\right>\left< \sigma_{1}^{+} \right>\left< a \right>-\left<
		\sigma_{1}^{-}\sigma_{2}^{+} \right>_c\left< a^{\dagger } \right>-\right.\right.\\
		\shoveleft[1cm] \left.\left.\left<
		\sigma_{1}^{-}a^{\dagger }
		\right>_c\left< \sigma_{1}^{+} \right>-\left<
		\sigma_{1}^{+}a^{\dagger }
		\right>_c\left< \sigma_{1}^{-} \right>-\left< \sigma_{1}^{+} \right>\left<
		\sigma_{1}^{-}
		\right>\left< a^{\dagger } \right>\right)+\right.\\
		\shoveleft[1cm]\left. \left< \sigma_{1}^{z}\sigma_{2}^{z} \right>_c\left<
		a^{\dagger }
		\right>+2\left< \sigma_{1}^{z}a^{\dagger } \right>_c\left< \sigma_{1}^{z}
		\right>+\left< \sigma_{1}^{z} \right>\left< \sigma_{1}^{z} \right>\left< a^{\dagger }
		\right> \right]
	\end{multlined}
	\label{eq:szis+j}
\end{align}

%\begin{align}
%	\left< \sigma_{1}^{+}\sigma_{1}^{-} \right>=\frac{1+\left< \sigma_{1}^{z} \right>}{2}
%	\label{eq:s1+s1+short}
%\end{align}

\begin{align}
	\frac{\mathrm{d}}{\mathrm{d}t}\left( \left< \sigma_{1}^{z}\sigma_{2}^{z} \right>_c+\left< \sigma_{1}^{z}
	\right>\left< \sigma_{1}^{z} \right> \right)&=-4\mathrm{i}g\left(
	\left<\sigma_{1}^{z}\sigma_{2}^{+}\right>_c\left< a \right>+\left< \sigma_{1}^{z}a
	\right>_c\left< \sigma_{1}^{+} \right>+\left< \sigma_{1}^{+}a \right>_c\left<
	\sigma_{1}^{z}
	\right>+\left< a \right>\left< \sigma_{1}^{z} \right>\left< \sigma_{1}^{+}
	\right>\right.\notag\\
	&\left.-\left< \sigma_{1}^{-}\sigma_{2}^{z} \right>_c\left< a^{\dagger} \right>-\left<
	\sigma_{1}^{z}a^{\dagger} \right>_c\left< \sigma_{1}^{-} \right>-\left<
	\sigma_{1}^{-}a^{\dagger} \right>_c\left< \sigma_{1}^{z} \right>-\left< a^{\dagger}
	\right>\left< \sigma_{1}^{-} \right>\left< \sigma_{1}^{z} \right>
	\right)\notag\\
	&-4\gamma_{a}\left( \left< \sigma_{1}^{+}\sigma_{1}^{-} \right>\left<
	\sigma_{1}^{z} \right> +\left< \sigma_{1}^{+}\sigma_{2}^{z} \right>_c\left<
	\sigma_{1}^{-} \right>+\left< \sigma_{1}^{-}\sigma_{2}^{z} \right>_c\left<
	\sigma_{1}^{+} \right>\right)
	\label{eq:sz1sz2}
\end{align}
\end{widetext}

\section{Validity of the cumulant expansion}\label{app:validity}

The validity of the truncation of the expansion performed previously relies on the assumption
that higher-order cumulants are negligible. This can be checked in principle by truncating
at higher orders and comparing the results.  In general, it turns out that there is one
cumulant which requires more care: the correlation between the inversion and the photon
number $\left< a^{\dagger }a \sigma_{1}^{z} \right>$. In the regime where
$\frac{g_{\text{eff}}}{\kappa}> 1$ and $\frac{g_{\text{eff}}}{\gamma_{a}}>1$ holds, the
number of photons necessary to saturate the ensemble is low. Therefore small fluctuations
of the photon number can cause significant changes in the inversion \citep{Rice1994}. The
correlation between the photon number and the inversion $\left< a^{\dagger }a
\sigma_{1}^{z} \right>_{c}$ is therefore kept in our calculations. Hence, an expansion like
in Eq.~\eqref{eq:expand3} would not be advantageous because none of the terms could be
dropped. We therefore calculate the dynamical equation for $\left< a^{\dagger }a
\sigma_{1}^{z} \right>$ which gives

\begin{figure}[!]
\centering
\subfigure{
\includegraphics[width=0.5\textwidth]{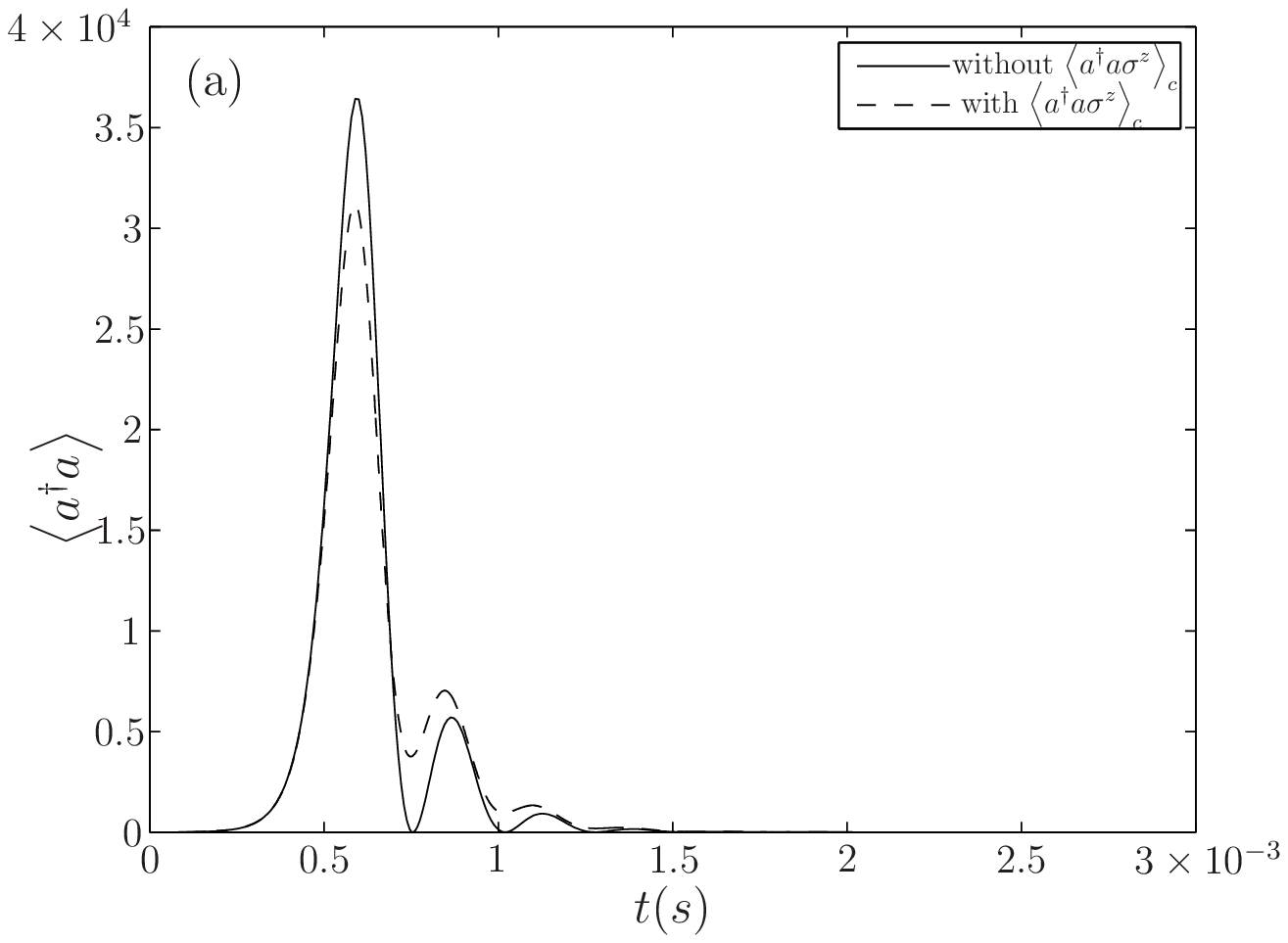}
}
\subfigure{
\includegraphics[width=0.5\textwidth]{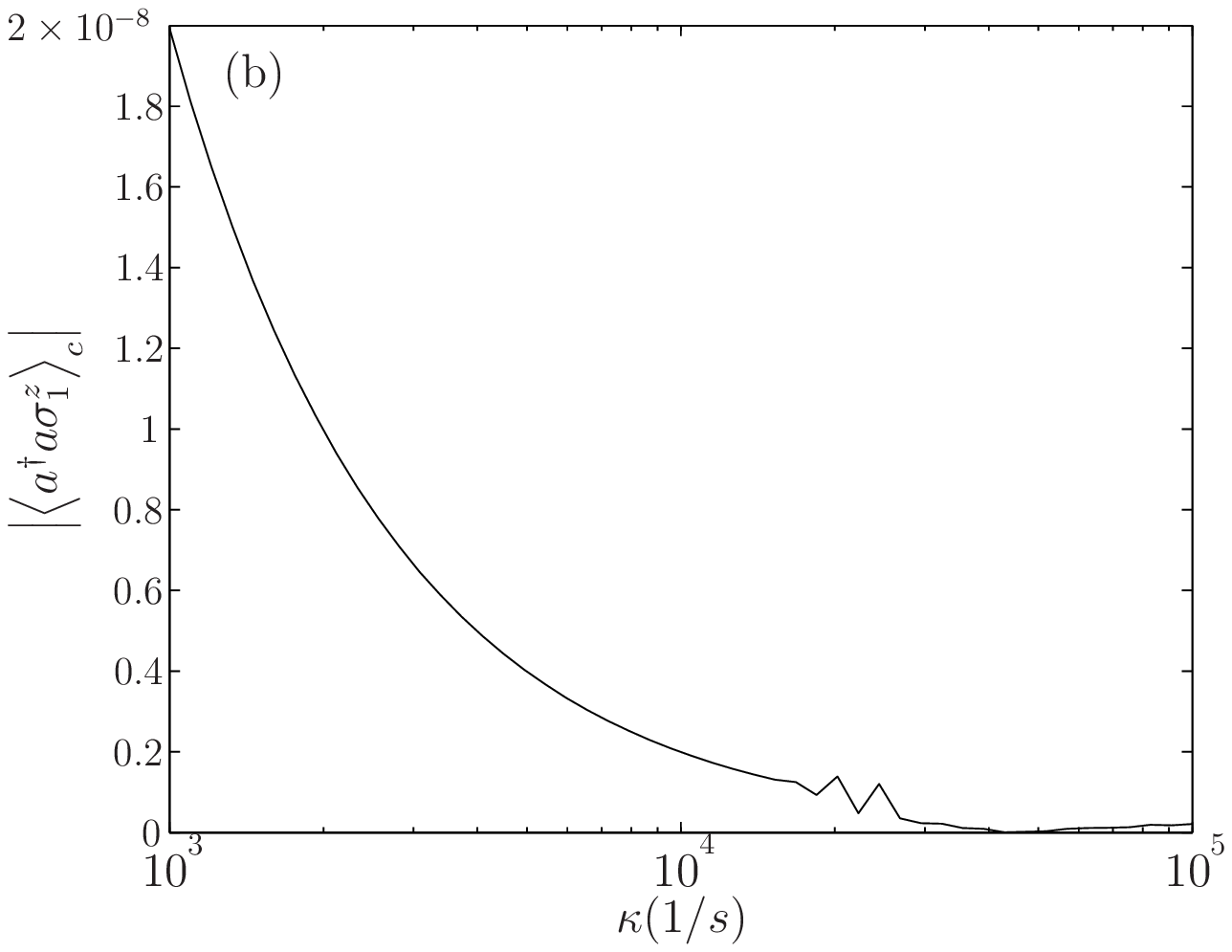}
}
\caption{$\left( \text{a} \right)$ Dynamics of the photon number in the cavity mode for an initially fully
	inverted ensemble. The solid line shows the dynamics produced by the set of equations where
	$\left< a^{\dagger }a\sigma_{1}^{z} \right>_{c}$ was neglected. The dashed curve is the
	result of the full set of 13 equations in which $\left< a^{\dagger
	}a\sigma_{1}^{z} \right>_{c}$ is kept. $\left( \text{b} \right)$ Numerically obtained cumulant $\left< a^{\dagger }a\sigma_{1}^{z} \right>_{c}$ in
the steady state. With increasing loss rate of the cavity $\kappa$ the correlation between
photon number and inversion decreases. }
\label{fig:get_the_cumulant_radiate_third_kappa10to3_10to5_g40_gamma03_T1_N10to5_eta10to3}
\end{figure}

\begin{widetext}
\begin{align}
	\frac{\mathrm{d}}{\mathrm{d}t}\left< a^{\dagger }a
	\sigma_{1}^{z} \right>=& \left(- 2\kappa-\gamma_{a} \right)\left< a^{\dagger }a
	\sigma_{1}^{z} \right> -\gamma_{a}\left< a^{\dagger }a \right> + 2\kappa\bar{n}\left<
	\sigma_{1}^{z}
	\right>\notag\\
	&-\mathrm{i} g \left( \left< a\sigma_{1}^{+} \right>-\left< a^{\dagger
	}\sigma_{1}^{-} \right>+2\left( \left< a^{\dagger }a a \sigma_{1}^{+} \right>-\left<
	a^{\dagger }a^{\dagger }a \sigma_{1}^{-} \right> \right) \right)\notag\\
	&-\mathrm{i}g\left( N-1 \right)\left( \left< a^{\dagger
	}\sigma_{1}^{z}\sigma_{2}^{-} \right>-\left< a\sigma_{1}^{z}\sigma_{2}^{+} \right>
	\right)\notag\\
	&+\eta\left< a^{\dagger }\sigma_{1}^{z} \right>+\eta^{*}\left< a\sigma_{1}^{z} \right>
	\label{eq:full_adasz}
\end{align}
\end{widetext}
The expectation values of products of three operators are expanded as in
Eq.~\eqref{eq:expand3}, except for $\left< a^{\dagger }a
\sigma_{1}^{z} \right>$. The expansion of expectation values with four operators is more
involved and produces also expectation values of products of three operators which are
again expanded. Cumulants of order three and four are neglected. The resulting equation for
$\left< a^{\dagger }a
\sigma_{1}^{z} \right>$ can
be integrated numerically along with the equations for the other quantities mentioned
previously.

To estimate the influence of the correlation between the photon number
and the inversion $\left< a^{\dagger }a\sigma_{1}^{z} \right>_{c}$ on the dynamics we plot the
photon number in the cavity during the decay of a fully inverted ensemble. We therefore
integrate a set of 12 equations that is obtained if $\left< a^{\dagger
}a\sigma_{1}^{z} \right>$ is expanded and $\left< a^{\dagger }a\sigma_{1}^{z} \right>_{c}$
is neglected. For comparison we also show the dynamics obtained from the full set of
13 equations in which $\left< a^{\dagger }a\sigma_{1}^{z} \right>_{c}$ is kept [see
Fig.~\ref{fig:get_the_cumulant_radiate_third_kappa10to3_10to5_g40_gamma03_T1_N10to5_eta10to3}
(a)].

The steady state of both solutions differs only slightly. The correlation $\left<
a^{\dagger }a\sigma_{1}^{z} \right>_{c}$ is shown in
Fig.~\ref{fig:get_the_cumulant_radiate_third_kappa10to3_10to5_g40_gamma03_T1_N10to5_eta10to3}
(b). With increasing cavity decay rate $\kappa$ the correlation between photon number and
inversion decreases.

\nocite{*}

\bibliography{aps_bib}

\end{document}